\documentclass[acmsmall]{acmart}
\AtBeginDocument{%
  \providecommand\BibTeX{{%
    \normalfont B\kern-0.5em{\scshape i\kern-0.25em b}\kern-0.8em\TeX}}}

\setcopyright{none}
\renewcommand\footnotetextcopyrightpermission[1]{} 
\acmISBN{978-1-4503-XXXX-X/18/06}

\usepackage{amsmath,amsfonts}
\usepackage{algorithmic}
\usepackage{subcaption}
\usepackage{graphicx}
\usepackage{verbatim}
\usepackage{multirow}
\usepackage{epigraph}
\usepackage{tcolorbox}
\usepackage{enumitem}
\usepackage{textcomp}
\usepackage{xcolor}
\usepackage{booktabs}
\usepackage{pifont}
\usepackage{mdframed}
\usepackage{listings}
\usepackage{url}
\usepackage{hyperref}
\usepackage{adjustbox}
\definecolor{LightGray}{gray}{0.9}

\def\BibTeX{{\rm B\kern-.05em{\sc i\kern-.025em b}\kern-.08em
    T\kern-.1667em\lower.7ex\hbox{E}\kern-.125emX}}

\begin{document}
\title{An Empirical Study of the Non-determinism of ChatGPT in Code Generation}

\author{Shuyin Ouyang}
\affiliation{%
  \institution{King's College London}
  \city{London}
  \country{United Kingdom}
  }

\author{Jie M. Zhang}
\affiliation{%
  \institution{King's College London}
  \city{London}
  \country{United Kingdom}
  }

\author{Mark Harman}
\affiliation{%
  \institution{University College London}
  \city{London}
  \country{United Kingdom}
  }

\author{Meng Wang}
\affiliation{%
  \institution{University of Bristol}
  \city{Bristol}
  \country{United Kingdom}
  }

\settopmatter{printacmref=false}

\begin{abstract}

There has been a recent explosion of research on Large Language Models (LLMs) for software engineering tasks, in particular code generation.
However, results from LLMs can be highly unstable; nondeterministically returning very different code for the same prompt.
Such non-determinism affects the correctness and consistency of the generated code, undermines developers' trust in LLMs, and yields low reproducibility in LLM-based papers. 
Nevertheless, there is no work investigating how serious this non-determinism threat is.

To fill this gap, this paper conducts an empirical study on the non-determinism of ChatGPT in code generation.
We chose to study ChatGPT because it is already highly prevalent in the code generation research literature. 
We report results from a study of 829 code generation problems across three code generation benchmarks (i.e., CodeContests, APPS, and HumanEval) with three aspects of code similarities: semantic similarity, syntactic similarity, and structural similarity.
Our results reveal that ChatGPT exhibits a high degree of non-determinism under the default setting: 
the ratio of coding tasks with zero equal test output across different requests is 75.76\%, 51.00\%, and 47.56\% for three different code generation datasets (i.e., CodeContests, APPS, and HumanEval), respectively.
In addition, we find that setting the \textit{temperature} to 0 does not guarantee determinism in code generation, although it indeed brings less non-determinism than the default configuration (\textit{temperature}=1). 
In order to put LLM-based research on firmer scientific foundations, researchers need to take into account non-determinism in drawing their conclusions.

\end{abstract}

\maketitle

\section{Introduction}

Large Language Models (LLMs) are nondeterministic by nature \cite{lee2022coauthor}.
This is because LLMs predict the probability of a word or token given the context, represented by a sample of words.
The randomness in LLMs typically comes from the sampling methods used during text generation, such as top-k sampling or nucleus sampling \cite{mitchell2023detectgpt, krishna2022rankgen}. 
As a result, 
identical instructions or prompts can yield completely different responses to separate requests.

This non-determinism (i.e., the inconsistency in the code candidates generated in different requests with identical prompts)\footnote{There are other terms in the literature that also refer to non-determinism, such as inconsistency, variance, randomness, and instability.}
is an essential consideration when using LLM in practice \cite{soares2022your}.
Unreliable and inconsistent code snippets can have significant negative effects on the process of software development, particularly in safety-critical applications where consistency and reliability are paramount \cite{kiviriga2023efficient, chatterjee2022reliability}.
It may also undermine developers' trust in LLMs when completely different suggestions are given at different times \cite{vaithilingam2022expectation}.

Moreover, non-determinism 
affects the reliability and reproducibility of empirical software engineering \cite{poesia2022synchromesh}.
Indeed, compared to other tasks of ChatGPT, such as question answering and text summarization, the non-determinism threat in code-related tasks is much more serious,
because the inconsistency (especially semantic inconsistency) often indicates errors in the generated code \cite{inala2022fault}.
It is therefore of vital importance to understand how serious the non-determinism is for LLM-based software engineering tasks and call for actionable solutions to alleviate this issue. 


This paper presents the first systematic empirical study on the threat of non-determinism of ChatGPT in code generation tasks.
We choose the code generation tasks because 
code generation with Large Language Models (LLMs), such as ChatGPT, has recently attracted significant attention due to its impressive and cutting-edge performance \cite{li2022competition, bubeck2023sparks, fan2023large}.
Indeed, many publications have emerged from 
both the software engineering community and the machine learning community on evaluating the capability of ChatGPT in code generation \cite{bubeck2023sparks, liu2023your, feng2023investigating, yeticstiren2023evaluating, bang2023multitask}.

This paper focuses on ChatGPT (including GPT-3.5 and GPT-4), rather than other LLMs, for the following two reasons:
1) ChatGPT is the most widely adopted LLM in code generation in the literature~\cite{fan2023large, liu2023your, feng2023investigating, liu2024refining, yu2024codereval, guo2024exploring, wang2024software};
2) ChatGPT has the best performance in code generation and represents the state-of-the-art so far~\cite{fan2023large, openai2023gpt4}.
Thus, as the first work on the non-determinism of LLMs in software engineering tasks, we focus on ChatGPT in this paper but encourage other work to continue to investigate the non-determinism issue in other LLMs.

We conduct a series of experiments using the ChatGPT models on three widely-studied code generation benchmarks (i.e. CodeContests, APPS, and HumanEval) with 829 coding problems. 
For each code generation task, we let ChatGPT make five predictions.
We then compare the similarity of the five code candidates from three aspects, namely \emph{semantic similarity}, \emph{syntactic similarity}, and \emph{structural similarity}.
We also explore the influence of temperature (i.e., a parameter that controls the randomness of the response generated by ChatGPT) on non-determinism, as well as the correlation between non-determinism
and coding task features such as the length of coding instruction and
the difficulty of the task.
We show the non-determinism with different models of ChatGPT, namely, GPT-3.5 and GPT-4.
Finally, we compare the non-determinism of code generation with different prompt engineering strategies.

Our results reveal that 
the threat of non-determinism in ChatGPT for code generation is serious, especially under default setting:
In particular, 
1) the ratio of problems with not a single equal test output among the top-five code candidates is above 50\% for all the benchmarks we study; 
2) the maximum difference of the test pass rate reaches 1.00 for all three datasets, and accounts for 39.63\% of the problems in HumanEval, the most widely used code generation benchmark;
In addition, 
contrary to the widely held belief (and practice followed to minimize nondeterminism)~\cite{deng2023large,bhavya2022analogy,liang2023code},
setting the temperature to zero
does not guarantee determinism in code generation.
Also interestingly, our result analysis suggests that 
the length of coding instructions has a negative correlation with almost all our similarity measurements, meaning that longer description length tends to yield code candidates with less similarity and more buggy code.
Different prompt engineering strategies also yield different degrees of non-determinism in code generation.

To understand how the literature handles the non-determinism threat, we collect 76 LLM-based code generation papers that appeared in the last 2 years.
Our manual analysis results highlight that only 21.1\% of these papers consider the non-determinism threat in their experiments. 
These results highlight that there is currently a significant threat to the validity of scientific conclusions.
We call for researchers to take into account the non-determinism threat in drawing their conclusions.

To summarize, this paper makes the following contributions:

\begin{itemize}[leftmargin=*]
    \item We present the first study of the non-determinism threat in code generation tasks on ChatGPT, with three widely-studied datasets (CodeContest, APPS, HumanEval) and three types of similarity measurements. Our results reveal that the non-determinism threat is serious and deserves attention from both academia and industry.
    \item We study the influence of temperature on the non-determinism of ChatGPT and find that setting temperature to zero does not guarantee determinism in code generation, which is contrary to many people's beliefs.
    \item We study the correlation between coding task features and the degree of non-determinism. The results reveal that the length of coding instruction has a negative correlation with syntactic and structural similarity, as well as the average correctness of the generated code.
    \item We study the influence of different prompt engineering techniques on code generation non-determinism. We find that prompts with a Chain-of-Thought strategy leads to more non-determinism when temperature=0, while code candidates generated from prompts requesting simple and concise code are more stable.

\end{itemize}

We release our data, code, and results at our homepage~\cite{homepage}.
The rest of the paper is organized as follows.
Section 2 introduces the main procedure of our study.
Section 3 describes the design of the experiments, including research questions, benchmarks, selected models, and measurement tools.
Section 4 presents the results and discusses some interesting findings based on the experimental results we obtained.
Section 5 discusses the threats to validity in two aspects, as well as the limitations of this study.
Section 6 introduces the related work of our study.
Section 7 discusses the implications for software developers and researchers and future work.
Section 8 concludes.

\section{Method}
\label{sec:promptpreparation}

\begin{figure}[htbp]
\centerline{\includegraphics[width=12cm]{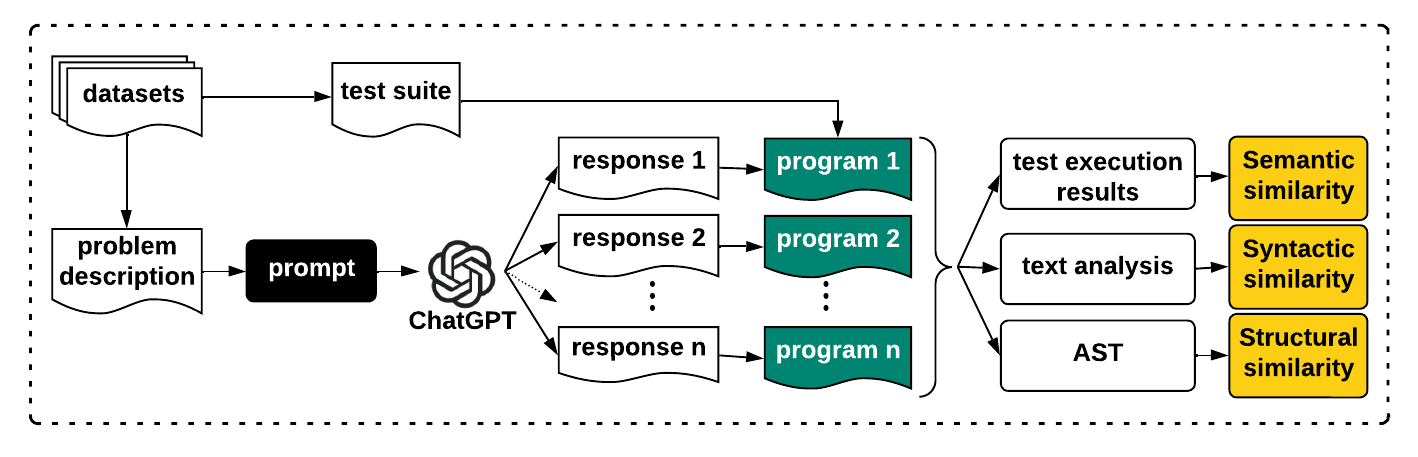}}
\vspace{0mm}
\caption{Overview of the experimental procedure.}
\label{fig: overview}
\end{figure}

Fig~\ref{fig: overview} shows an overview of our experimental procedure. 
For each code generation task,
our study first produces a prompt with a coding instruction,
then feeds this prompt to ChatGPT API~\cite{openaichat} to generate code (zero-shot).
We call the API five times to let ChatGPT make five predictions with the same prompt. 
We then extract code from each of the five responses, 
to get five code candidates.
Our non-determinism analysis compares the five code candidates in terms of their semantic similarity, syntactic similarity, and structural similarity. 

\noindent\textbf{Prompt synthesis}: The first step in our study is prompt preparation.
There are many ways to conduct prompt engineering for code generation.
In this paper, we follow the common practice in LLM-based code generation assessment~\cite{austin2021program,fan2023large}.
In particular,
1) we ask ChatGPT to generate Python code for each code generation task with zero-shot prompting;
2) we use the basic prompt design directly followed by programming task descriptions.
To guarantee that ChatGPT produces code rather than pure natural languages in its response,
we augment the original coding problem description with an instruction to request for Python code.

One challenge in extracting the code from the API response is that there is no clear signal to distinguish code with plain text in the response, which is different from ChatGPT's web chat window (i.e. in the chat window, codes are returned with Markdown code blocks). 
To address this problem, we specify the format of the generated code into `Markdown'. 
Thus, for each code generation task, our prompt is shown as follows:


\begin{lstlisting}
# this is the original coding problem description.
# Generate Python3 code (Markdown):
\end{lstlisting}

\noindent\textbf{Code Extraction}:
After receiving the response from ChatGPT, we apply code extraction to retrieve the code from the generated text.
We compile the code directly without making any modifications. Our experiments are mainly run on Google Deep Learning VM instances, with the Linux environment pre-installed from open images\footnote{\url{https://cloud.google.com/compute/docs/images}}. All of the necessary libraries are pre-installed. In this way, it can ensure to the greatest extent that the generated code will not cause import errors caused by the library not being installed during running.

\noindent\textbf{Test Case Execution}:
To evaluate the semantics of ChatGPT's generated code, we use the test suite that is suited to each benchmark.
We not only record whether each test passes or not but also record every specific test output, which enables us to compare the similarity of test outputs even if they both fail.
For CodeContests and HumanEval datasets, every problem has a certain timeout value of 3 seconds. The APPS dataset does not provide a default timeout value, and we set the value to be 3 seconds as well.
We use single-threaded scripts to run the tests to ensure that the test cases are executed sequentially to avoid race conditions that may arise from concurrent executions.

\noindent\textbf{Similarity Checking}:
To measure the similarity between code candidates, we introduce similarity measurement tools that evaluated the semantic, syntactic, and structural similarity between the generated code solutions. The semantic similarity is measured by comparing test execution outputs. The syntactic similarity is measured by comparing the text similarity between codes. The structural similarity is evaluated by comparing the code candidates' abstract syntax trees (ASTs).
More details about our similarity measurement methods are mentioned in Section~\ref{subsec:similaritymeasurement}.

\section{Experimental Design}

\subsection{Research Questions}

This study answers the following questions:

\noindent \textbf{\emph{RQ1: To what extent is ChatGPT susceptible to non-determinism in code generation under the default setting?}} This RQ investigates the non-determinism of ChatGPT in terms of the semantic, syntactic, and structural similarity among the code candidates generated with identical instructions under the default setting.
There are three sub-RQs:
\begin{itemize}[leftmargin=6mm]
    \item{\emph{Sub-RQ1.1: To what extent is ChatGPT susceptible to non-determinism in terms of semantic similarity?}}
    \item{\emph{Sub-RQ1.2: To what extent is ChatGPT susceptible to non-determinism in terms of syntactic similarity?}}
    \item{\emph{Sub-RQ1.3: To what extent is ChatGPT susceptible to non-determinism in terms of structural similarity?}}
\end{itemize}

\noindent \textbf{\emph{RQ2: How does temperature affect the degree of non-determinism?}}
Temperature is a hyperparameter of LLMs for controlling the randomness of the predictions.
This RQ checks and compares the non-determinism of ChatGPT in code generation with different choices of temperature.

\noindent \textbf{\emph{RQ3: How does the non-determinism compare to the similarity of the top code candidates generated within the same prediction?}} 
ChatGPT can be configured to generate multiple candidates for one prediction, which are ranked by their predictive probability.
This RQ compares the similarity of the code candidates obtained in different predictions with those obtained within the same prediction.

\noindent \textbf{\emph{RQ4: What types of coding tasks have a higher degree of non-determinism?}} To understand what affects non-determinism, this RQ studies the correlation between the features of coding tasks (e.g., the length of code generation instructions, the code problem difficulty, and labels) and the similarity metrics used in our study.
We also conduct qualitative analysis on specific cases for deep analysis. 

\noindent \textbf{\emph{RQ5: How is GPT-4's non-determinism compared with GPT-3.5?}} This RQ compares GPT-3.5 and GPT-4 in their degree of non-determinism in generating code. 

\noindent \textbf{\emph{RQ6: How do different prompt engineering strategies influence the degree of non-determinism?}} This RQ compares the degree of non-determinism for different prompt engineering strategies (i.e., Chain-Of-Thought and requesting generated code as concise as possible) when using ChatGPT to generate code.

\subsection{Code Generation Benchmarks}

Our experiments use the three most widely studied code generation benchmarks: CodeContest \cite{li2022competition}, APPS \cite{hendrycks2021measuring}, and HumanEval \cite{chen2021evaluating}. 
Table~\ref{tab:benchmarks} shows their details.
Each of these datasets has unique characteristics, which are introduced below.
The distribution of difficulty and problem tags of these datasets are available on our homepage~\cite{homepage}.

\begin{table}[htbp]\scriptsize
\caption{\label{tab:benchmarks}Code generation benchmarks}
\vspace{0mm}
\begin{center}
\begin{tabular}{l r r r r}
\toprule
Name & Mean Length & No. of & Mean No. of & Mean No. of Provided \\
& of description & Problems & Test Cases &  Correct Solutions\\
\midrule
CodeContests & 1989.19 & 165 & 203.84 & 49.99 \\
APPS & 1663.94 & 500 & 80.43 & 20.92 \\
HumanEval & 450.60 & 164 & 9.24 & 1.00\\
\bottomrule
\end{tabular}
\label{tab1}
\end{center}
\end{table}

\noindent \textbf{CodeContests}: CodeContests is used when training AlphaCode, which comprises coding problems from various sources such as Aizu\footnote{\url{https://judge.u-aizu.ac.jp}}, AtCoder\footnote{\url{https://atcoder.jp}}, CodeChef\footnote{\url{https://www.codechef.com}}, CodeforcesCodeChef\footnote{\url{https://codeforces.com}}, and HackerEarthCodeChef\footnote{\url{https://www.hackerearth.com}}. In our experiment, following the assessment practice of 
AlphaCode,
we use the test set of CodeContests to benchmark the code generation tasks of ChatGPT.

\noindent \textbf{APPS}: APPS includes 10,000 coding problems (both the training set and testing set).
This dataset is exclusively designed for Python program synthesis evaluation. 
The original test set contains 5,000 code-generation problems, and we 
randomly sample 500 problems, among which there are 60.20\% interview problems, 19.60\% introductory problems, and 20.20\% competition problems.
APPS evaluates models not only on their ability to code syntactically correct programs but also on their ability to understand task descriptions and devise algorithms to solve these tasks~\cite{hendrycks2020aligning}.

\noindent \textbf{HumanEval}: The HumanEval dataset is an evaluation set first proposed in \cite{chen2021evaluating}, which contains 164 hand-written coding problems. Each problem includes a function signature, docstring, body, and several unit tests, with an average of 9.24 test cases per problem. We use the whole dataset to benchmark our experiments.

As mentioned in Section~\ref{sec:promptpreparation}, we especially focus on the code generated with Python3 language, since it is one of the most widely studied programming languages in code generation~\cite{wei2019code, sun2019grammar, li2022competition, chen2021evaluating, feng2020codebert, austin2021program, svyatkovskiy2020intellicode}.

\subsection{Configuration of ChatGPT}

ChatGPT has gained widespread popularity and recognition in multiple tasks including question-answering, language translation, sentiment analysis, and text summarising, among which code generation is one of the most impressive tasks \cite{li2022competition, bubeck2023sparks}.
There are several reasons why we have chosen ChatGPT as our research target among all large language models.
Firstly, ChatGPT has the ability to generate highly coherent and contextually appropriate responses to a wide variety of textual prompts  \cite{hassani2023role}. This makes it an ideal tool for conducting research in areas of code generation by designing specific prompts. Secondly, the GPT-3.5 series is a particularly attractive option due to its impressive performance and large-scale training data, which allows for more accurate and nuanced language processing capabilities \cite{liu2023summary}. 
Thirdly, the model API `gpt-3.5-turbo' and `gpt-4' released with ChatGPT have not been extensively studied in academia, and their capabilities in terms of code generation are thus still unknown.
Therefore, we choose them as our experiment target models. 
Written in ChatGPT's official website\footnote{\url{https://platform.openai.com/docs/api-reference/chat/create}}, using ChatGPT’s model API requires various parameters. We use the default values for most of the parameters in addition to the following ones:

\begin{itemize}[leftmargin=*]
  \item \textbf{\textit{model}}: ID of the model to use. This parameter is strictly required, and in our case, we set this parameter to `gpt-3.5-turbo-0125' or `gpt-4-0613'. 

  \item \textbf{\textit{message}}: A list of messages describing the conversation so far, where two key values `role' and `content' should be filled. This parameter is also strictly required. In our experiments, the message's `role' is `user' and the `content' contains the prompt we used for requesting for all of the RQs.
  
  \item \textbf{\textit{temperature}}: What sampling temperature to use, between 0 and 2 (Default value is 1). Higher values will make the output more random, while lower values will make it more focused and deterministic. 
  In our study, we study the influence of temperature in RQ2 with three temperature values: 0, 1, and 1.5. For RQ1, we use \textit{temperature}=1 only, and for the rest of the RQs, we present results with both \textit{temperature}=1 and \textit{temperature}=0.
  \item \textbf{\textit{top\_p}}: An alternative to sampling with temperature, called nucleus sampling, where the model considers the results of the tokens with \textit{top\_p} probability mass. In our experiment, we do not take it into consideration and set this value to remain at its default setting (i.e., \textit{top\_p}=1).
  \item \textbf{\textit{n}}: How many code candidates (the so-called ``chat completion choices'' according to the ChatGPT API website~\cite{openaichat}) to generate for each input message (with 1 being the default value). The default value of \textit{n} is 1. In RQ3, we set \textit{n}=5 to investigate how the non-determinism of code candidates from the same request compares with those from different requests. We choose \textit{n}=5, since 5 is a widely used figure in the papers studying variance~\cite{nagarajan2018deterministic}. 
  \textit{n}=5 is only used in RQ3. 
\end{itemize}

\subsection{Non-determinism Measurement}
\label{subsec:similaritymeasurement}

In order to answer our research questions, we introduce the following tools for measuring the degree of non-determinism.

\subsubsection{Semantic similarity} We measure the semantic similarity of different code candidates by checking their similarity in test execution results, including \textbf{\emph{test pass rate}} and \textbf{\emph{output equivalence rate}}.
The test pass rate calculates the ratio of the passed test case number against the total test case number for code candidates.
It is one of the most widely used measurement metrics for assessing code generation capabilities\footnote{Although the benchmarks are very widely studied, their test suites can be inadequate. This paper is less affected by the inadequate test suite issue as we focus on the similarity of test pass rate, rather than the absolute value of test pass rate.}~\cite{zan2022cert, li2022competition, chen2021evaluating, hendrycks2021measuring, austin2021program}.
Each code generation problem has five test pass rates, one for each code candidate. 
We use the variance and maximum difference of the five values to indicate semantic similarity.
We also calculate the mean of the five values for the purpose of understanding correctness as well as the correlation between correctness and non-determinism (RQ4).

The output equivalence rate records the ratio of identical test outputs (across different code candidates for the same code generation instruction) against the total test outputs.
Each instruction has one output equivalence rate.
For tests that produce specific outputs (without exceptions or errors), we check whether the output values of different code candidates are equal to each other. 
In the following parts of this paper, we use OER to represent output equivalence rate and use OER (no ex.) to represent output equivalence rate (without exceptions or errors) for short.
Each code generation problem has only one OER and OER (no ex.).
Additionally, we measure the OER and OER (no ex.) in pairs and report the mean output equivalence rate of the combinations of every two code candidates for a coding problem.
For tests that yield exceptions or timeout errors, we consider the test outputs to be the same if the exception or error messages are the same.

Some papers use the pass@k metric~\cite{kulal2019spoc,chen2021evaluating} (i.e., the ratio of coding tasks with 100\% test pass rate) to indicate the high-level code generation correctness of a code generation approach. We do not use this metric in our main body of experiments because we focus on the non-determinism threat, while pass@k ignores the correctness of each single coding task and concentrates only on the ratio of correct code candidates in all the tasks, which can cover the non-determinism across different requests.
In addition, pass@k does not reflect the practical application scenario of LLMs in code generation, because developers are less likely to 
try the model for k times until they finally get one correct solution.


\subsubsection{Syntactic similarity} 

The syntactic similarity in this study treats different code candidates as texts and checks their textual similarity.
We choose the \textbf{Longest Common Subsequence} and \textbf{Levenshtein Edit Distance} as evaluation tools~\cite{li2023skcoder, mastropaolo2023robustness, li2023codeeditor, xu2022ide}. 
In the following content, we use LCS and LED to represent the Longest Common Subsequence and Levenshtein Edit Distance for short respectively.
LCS measures the similarity via the normalized length of the longest common subsequence between two sequences.
LED measures the minimum number of single-token edits (insertions, deletions, or substitutions) required to change one code into the other.
LCS and LED both regard the token as the smallest unit, and the token is divided by the \texttt{.split()} method, that is, any whitespace is used as the separator to divide the code into tokens.
We measure the syntactic similarity with LCS/LED by comparing the first code candidate with each of the remaining four code candidates. 
Thus, each code-generation problem has four values of each metric. 
We use the mean, mean worst value (i.e., mean highest value for LED and mean lowest value for LCS), and pair mean (by comparing all the combinations of two code candidates in pairs) to indicate the syntactic similarity measured by each metric.

Below are the formulas for the LCS and LED:

\begin{equation*}
LCS = \frac{len(lcs(s, t))}{len(s)}
\end{equation*}
where \(s\) is reference string, \(t\) is the string to be compared, \(lcs(s,t)\) is the longest common subsequence between \(s\) and \(t\).


\[
{\text{LED}_{s,t}(i,j) =
\begin{cases}
\max(i,j) & \text{if } \min(i,j) = 0 \\
\min
\begin{cases}
\text{led}_{s,t}(i-1,j) + 1 \\
\text{led}_{s,t}(i,j-1) + 1 \\
\text{led}_{s,t}(i-1,j-1) + 1_{(s_i \neq t_j)}
\end{cases} & \text{otherwise}
\end{cases}}
\]
where $\operatorname{LED}_{s,t}(i,j)$ is the LED between the first $i$ characters of $s$ and the first $j$ characters of $t$, and $\operatorname{diff}(s_i,t_j)$ is $0$ if the $i$-th character of $s$ is the same as the $j$-th character of $t$, and $1$ otherwise.

\subsubsection{Structural similarity} 
We design structural similarity to measure the code similarity in terms of the Abstract Syntax Tree (AST). 
AST is a tree-like representation of the source code in which each node in the tree represents a construct in the code, such as variable, function, or control structure, and the edges between nodes represent the relationships between these constructs.
We use a Python library called \texttt{pycode\_similar}\footnote{\url{https://github.com/fyrestone/pycode_similar}} \cite{li2023cctest, wu2020deep} to calculate the similarity. 
The \texttt{pycode\_similar} normalizes Python code into AST representation and uses Python library \texttt{difflib} to get the modification from referenced code to target code. There are two different measurement settings, i.e. \textbf{Unified\_Diff} and \textbf{Tree\_Diff}. 
Unified\_Diff measures the difference of normalized function AST string lines, while Tree\_Diff measures the difference in tree edit distance between two given ASTs.
Similar to syntactic similarity,
for each code generation problem, we report the mean, smallest similarity values, and pair mean among the five candidates.

\subsubsection{Statistical Analysis}

We conduct statistical analysis to demonstrate the significance of the differences among the outputs. 
We choose Kruskal-Wallis test~\cite{mckight2010kruskal} which does not require assumptions of normal distribution. 
The Kruskal-Wallis test stands as a non-parametric method for analyzing data, serving as an extension of the Mann-Whitney U test~\cite{mcknight2010mann} to more than two independent groups.
The essence of the Kruskal-Wallis test lies in comparing the median ranks among groups, rather than the means, which makes it robust against outliers and non-normal distribution of data.

\section{Results and Findings}
This section introduces the experimental results as well as the analysis and discussion for each RQ.

\subsection{RQ1: Non-determinism of ChatGPT with Three Types of Similarities under default setting}

\subsubsection{RQ1.1: Semantic Similarity}
\begin{figure}[h!]
  \centering
  \begin{subfigure}{0.3\linewidth}
    \centering
    \includegraphics[width=\linewidth]{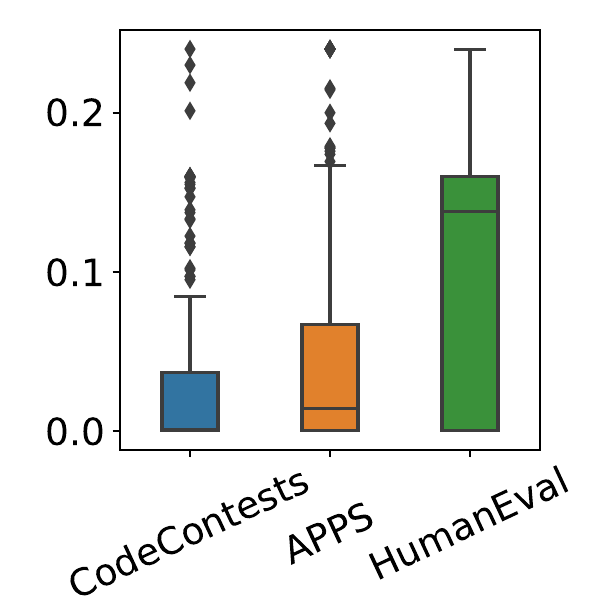}
    \caption{Variance}
  \end{subfigure}
  \hfill
  \begin{subfigure}{0.3\linewidth}
    \centering
    \includegraphics[width=\linewidth]{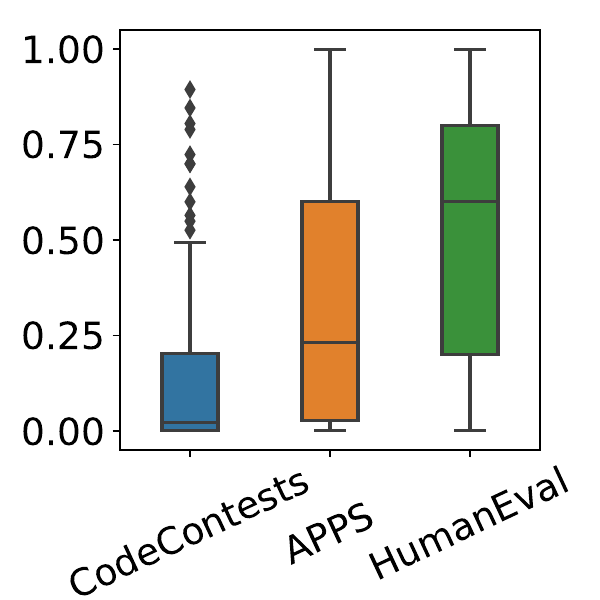}
    \caption{Mean}
  \end{subfigure}
  \hfill
  \begin{subfigure}{0.3\linewidth}
    \centering
    \includegraphics[width=\linewidth]{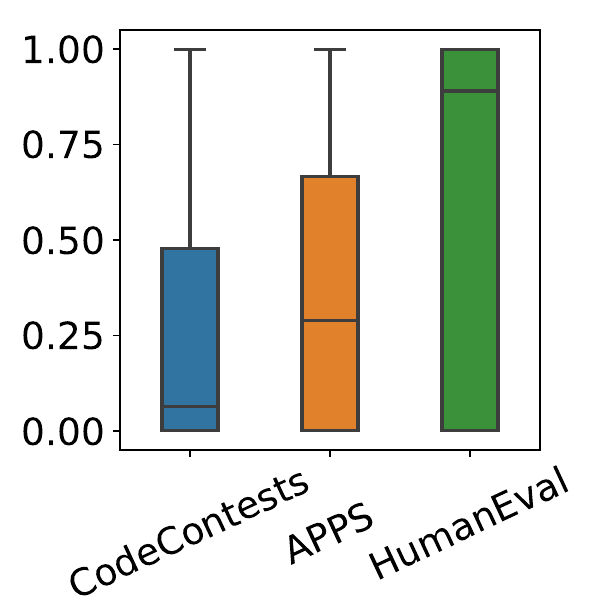}
    \caption{Max Diff}
  \end{subfigure}
  \vspace{0mm}
  \caption{RQ1.1: Distribution of semantic similarity in terms of test pass rate.}
  \label{fig:test pass rate}
\end{figure}

\begin{figure}[h!]
  \centering
  \begin{adjustbox}{width=0.6\textwidth}
  \begin{subfigure}{0.45\linewidth}
    \centering
    \includegraphics[width=\linewidth]{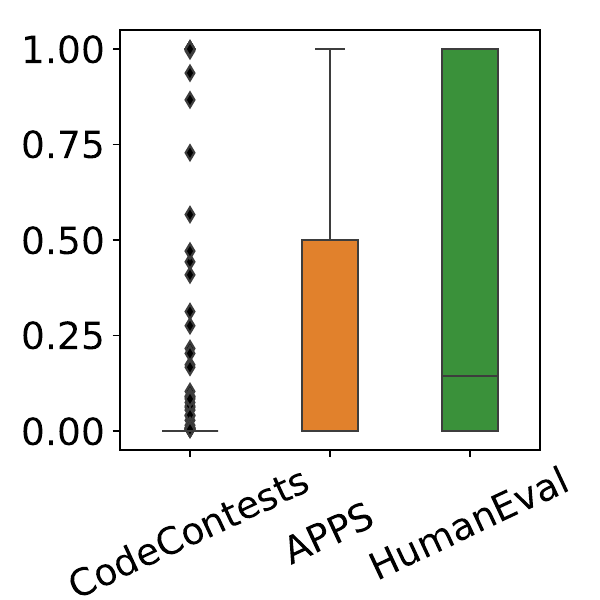}
    \caption{OER}
  \end{subfigure}
  \hfill
  \begin{subfigure}{0.45\linewidth}
    \centering
    \includegraphics[width=\linewidth]{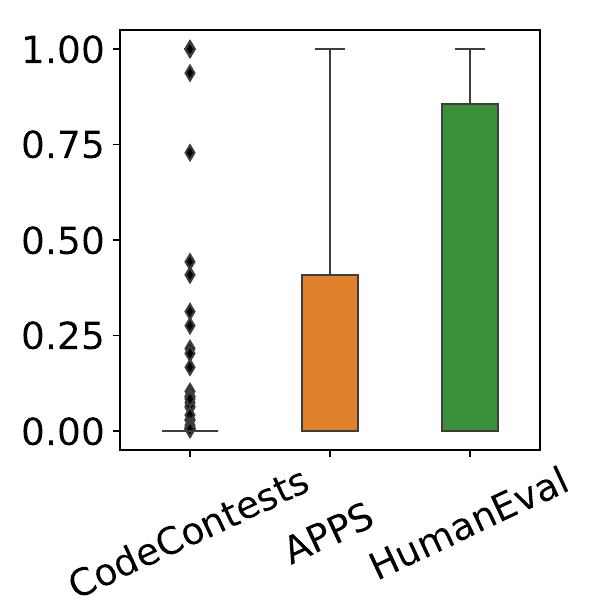}
    \caption{OER (no ex.)}
  \end{subfigure}
   \end{adjustbox}
   \vspace{0mm}
  \caption{RQ1.1: Distribution of semantic similarity in terms of test output equivalence rate (OER and OER (no ex.)).}
  \label{fig: OER and OER no ex.}
\end{figure}

Semantic similarity is measured by the following metrics: test pass rate and OER (output equivalence rate), and OER excluding exceptions.
As mentioned in Section~\ref{subsec:similaritymeasurement},
each coding problem has five test pass rates, we use the variance and maximum difference of these five values to indicate ChatGPT's non-determinism in generating code for the task.
We also report the mean value, which represents the average correctness of the generated code.
For OER or OER (no ex.), 
we compare the equivalence across all the five code candidates as well as between every two candidates. 
For each dataset, we report the distribution of different measurements in Figure~\ref{fig:test pass rate} and Figure~\ref{fig: OER and OER no ex.}.
The mean measurement values for all the coding problems (the mean value inside each bar in each bar chart) in a dataset are shown in Table~\ref{table: Output Features}.
The max diff refers to the maximum value of the max diff among all the coding problems.
In addition, Table~\ref{table: Output Features} also shows the ``Ratio of worst cases'', which is the ratio of problems with maximum diff of test pass rate being 1 or OER being 0.

\begin{table}[h!]\scriptsize
\caption{RQ1.1: Results of semantic similarity. OER and OER (no ex.) are the output equivalence rate and the equivalence rate excluding exceptions.}
\vspace{0mm}
\begin{tabular}{l l r r r}
\toprule
Semantic similarity & Metric & CodeContests  & APPS & HumanEval\\
\midrule
\multirow{5}{*}{Test pass rate} & Mean value & 0.16 & 0.42 & 0.63\\
&  Mean variance & 0.03 & 0.04 & 0.09 \\
&  Mean max diff & 0.24 & 0.35 & 0.53 \\
&  Max diff & 1.00 & 1.00 & 1.00  \\
&  Ratio of worst cases & 3.64\%  & 10.40\% & 39.63\%\\
\midrule
\multirow{3}{*}{OER} & Mean value &  0.09 & 0.27 & 0.39\\
&  Pair mean value & 0.27 & 0.47 & 0.67 \\
&  Worst value & 0.00 & 0.00 & 0.00 \\
&  Ratio of worst cases & 75.76\% & 51.00\% & 47.56\% \\
\midrule

\multirow{3}{*}{OER (no ex.)} & Mean value & 0.06 & 0.25 &  0.35 \\
&  Pair mean value & 0.19 & 0.42 & 0.61 \\
&  Worst value & 0.00 & 0.00 & 0.00 \\
&  Ratio of worst cases & 81.21\% & 53.40\% & 51.22\% \\

\bottomrule
\end{tabular}
\label{table: Output Features}
\end{table}



From Figure~\ref{fig:test pass rate}, Figure~\ref{fig: OER and OER no ex.}, and Table~\ref{table: Output Features},
we observe that ChatGPT is very unstable in generating semantically consistent code candidates. 
In particular, the ratios of tasks with zero equal test output (i.e., OER=0) among the five code candidates are 75.76\%, 51.00\%, and 47.56\% for the three datasets, respectively. 
This indicates that for the majority of the cases, ChatGPT generates code candidates with completely different semantics from identical instructions.

The mean variance of the test pass rate is relatively small from Table ~\ref{table: Output Features}, ranging between 0.03 and 0.09, this is because the test pass rate of different code candidates is often equally worse, as can be observed from Figure~\ref{fig:test pass rate}.(a).
However, the max diff of the test pass rate reaches 1.00 for all three datasets and accounts for 39.63\% of the problems in HumanEval, the most widely used code generation benchmark.
This indicates the correctness of code candidates generated from the same instruction can vary significantly. 
The large difference in different datasets also sheds light on the importance of using multiple datasets when assessing the code generation performance for large language models.

Our statistical analysis with Kruskal-Wallis test shows that, in 92.1\% of CodeContests, 39.4\% of APPS, and 40\% of HumanEval, the outputs of the code are indeed significantly different, where the p-value under the Kruskal-Wallis test is less than 0.05.


\begin{tcolorbox}
\textbf{\underline{Answer to RQ1.1:}}
The semantic difference among the code generated by ChatGPT in different requests is significant.
In particular, the ratio of coding tasks with not a single equal test output among the five different requests is 75.76\%, 51.00\%, and 47.56\% for CodeContests, APPS, and HumanEval, respectively. 
In addition, the maximum difference of the test pass rate reaches 1.00 for all three datasets and accounts for 39.63\% of the problems in HumanEval, the most widely used code generation benchmark. 
\end{tcolorbox}

\subsubsection{RQ1.2: Syntactic Similarity}

\begin{figure}[h!]
  \centering
  \begin{subfigure}{0.3\linewidth}
    \centering
    \includegraphics[width=\linewidth]{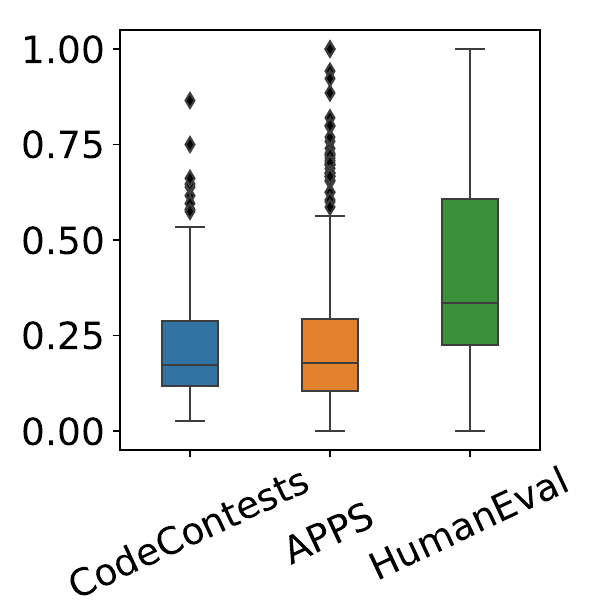}
    \caption{LCS mean}
  \end{subfigure}
  \hfill
  \begin{subfigure}{0.3\linewidth}
    \centering
    \includegraphics[width=\linewidth]{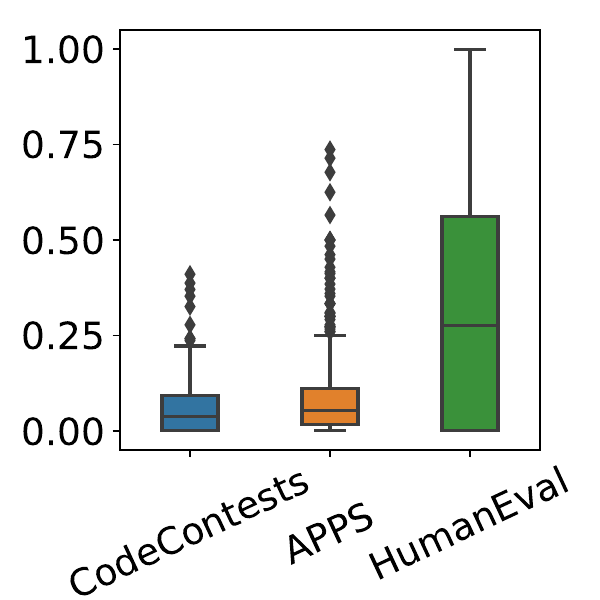}
    \caption{LCS Worst}
  \end{subfigure}
  \hfill
  \begin{subfigure}{0.3\linewidth}
    \centering
    \includegraphics[width=\linewidth]{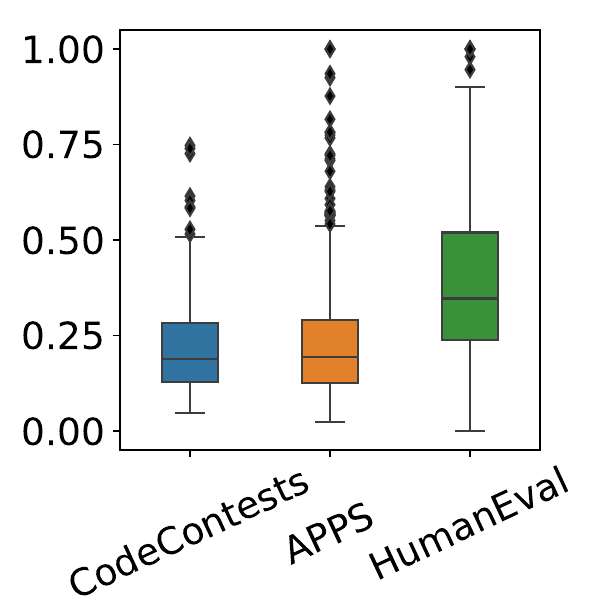}
    \caption{Pair LCS}
  \end{subfigure}

  \begin{subfigure}{0.3\linewidth}
    \centering
    \includegraphics[width=\linewidth]{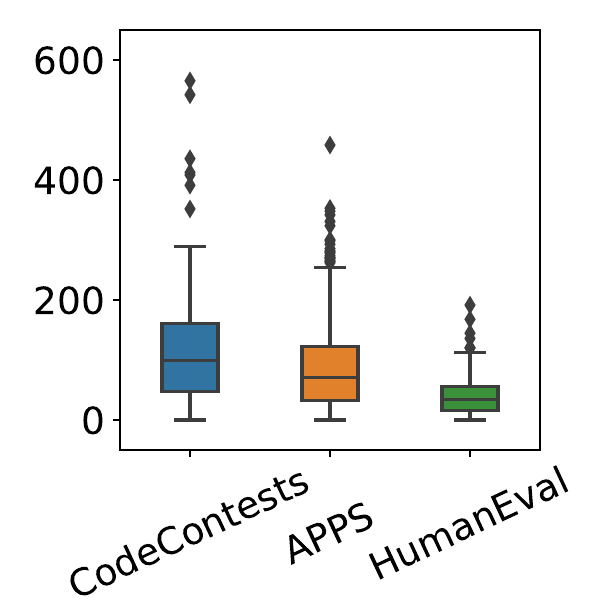}
    \caption{LED Mean}
  \end{subfigure}
  \hfill
  \begin{subfigure}{0.3\linewidth}
    \centering
    \includegraphics[width=\linewidth]{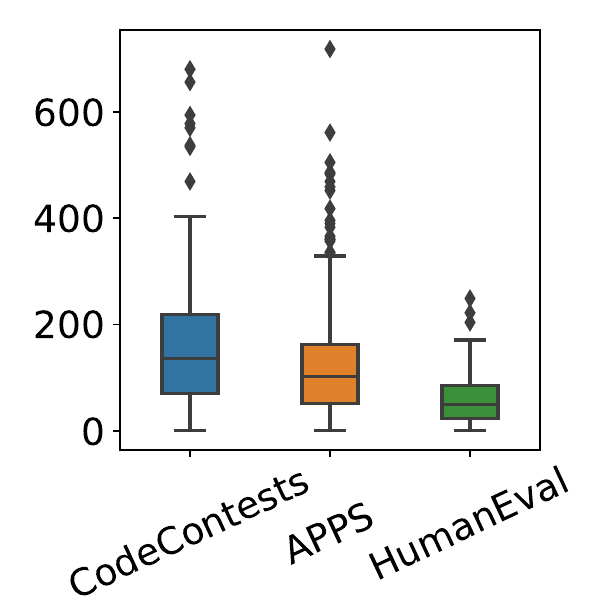}
    \caption{LED Worst}
  \end{subfigure}
  \hfill
  \begin{subfigure}{0.3\linewidth}
    \centering
    \includegraphics[width=\linewidth]{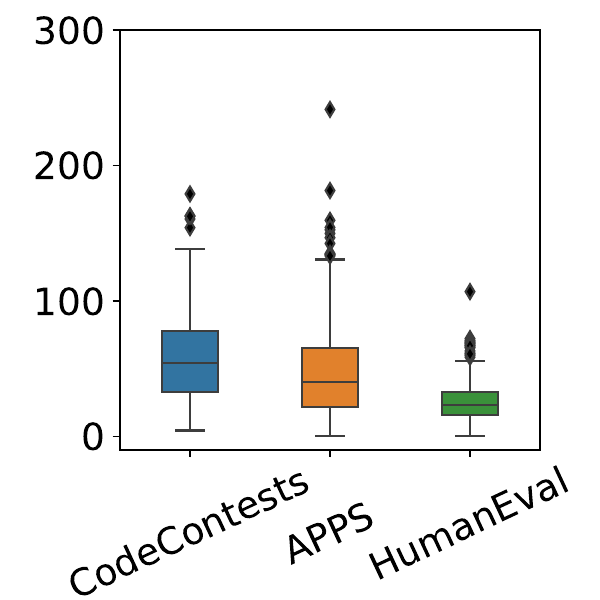}
    \caption{Pair LED}
  \end{subfigure}
  \vspace{0mm}
  \caption{RQ1.2: Distribution of syntactic similarity (LCS \& LED). Lower LCS and higher LED indicate less syntactic similarity.}
  \label{fig: syntacticsimilarity}
\end{figure}

Syntactic similarity measures the text similarity among code candidates.
In our experiment, the syntactic similarity is evaluated by the following metrics: LCS and LED (more details in Section~\ref{subsec:similaritymeasurement}).
For the five code candidates for each coding problem, we use the first code candidate as a reference and calculate the LCS and LED between the reference and the remaining four candidates. 
In addition, we calculate LCS and LED with code candidates in pairs, for each pair combination.
Thus, each problem has four LCS values and LED values, and 20 LCS and LED values in pairs, each value indicating a syntactic similarity.
We use the mean of these four values as well as the worst of them (i.e., the smallest value for LCS and the largest value for LED), and the mean of these 20 values calculated in pairs to represent each problem's syntactic similarity.
Figure \ref{fig: syntacticsimilarity} shows the distribution of LCS and LED for all the problems in each dataset.
Table~\ref{tab:syntacticsimilarity} shows the mean,  mean worst, and pair mean LCS and LED values for all the coding problems (the mean value inside each bar in the figures) in a dataset.



\begin{table}[h!]\scriptsize
\centering
\caption{RQ1.2: Syntactic similarity. Lower LCS and higher LED indicate lower syntactic similarity.}
\vspace{0mm}
\begin{tabular}{l l r r r}
\toprule
Syntactic Similarity & Metric & CodeContests & APPS & HumanEval\\
\midrule
\multirow{2}{*}{LCS} & Mean value & 0.22 & 0.23 & 0.42 \\
& Mean worst value & 0.16 & 0.16 & 0.25\\
&  Pair mean value &  0.23 & 0.24 & 0.41\\
\midrule
\multirow{2}{*}{LED} & Mean value & 58.80 & 47.37 & 26.56\\
& Mean worst value & 77.46 &  61.55 & 43.91\\
& Pair mean value & 58.86 & 46.94 & 27.10\\
\bottomrule

\end{tabular}
\label{tab:syntacticsimilarity}
\end{table}



We observe that the code candidates generated from the same instruction also differ largely in the syntactic measure.
Specifically,
the mean LCS is 0.22, 0.23, and 0.42 for CodeContests, APPS, and HumanEval, respectively,
indicating the mean ratio of the longest 
common subsequences among the code candidates.

For the three datasets, 
we could see from Table~\ref{tab:syntacticsimilarity} that the lowest LCS and largest LED values both happen for the CodeContests dataset. 
By contrast, the largest LCS and smallest LED values both happen for HumanEval.
This indicates that ChatGPT is most unstable syntactically for the code generation tasks in CodeContests, and most stable for HumanEval.
We further explore the correlation between different similarities and code task features in Section~\ref{subsec:correlationbetweencodetasks}.

\begin{tcolorbox}
\textbf{\underline{Answer to RQ1.2:}} 
Code candidates generated by ChatGPT in different requests also differ significantly in syntax. 
The mean syntax similarity (LCS) is only 0.22, 0.23, and 0.42 for CodeContests, APPS, and HumanEval, respectively.
\end{tcolorbox}

\subsubsection{RQ1.3: Structural Similarity}

\begin{figure}[h!]
  \centering
  \begin{subfigure}{0.3\linewidth}
    \centering
    \includegraphics[width=\linewidth]{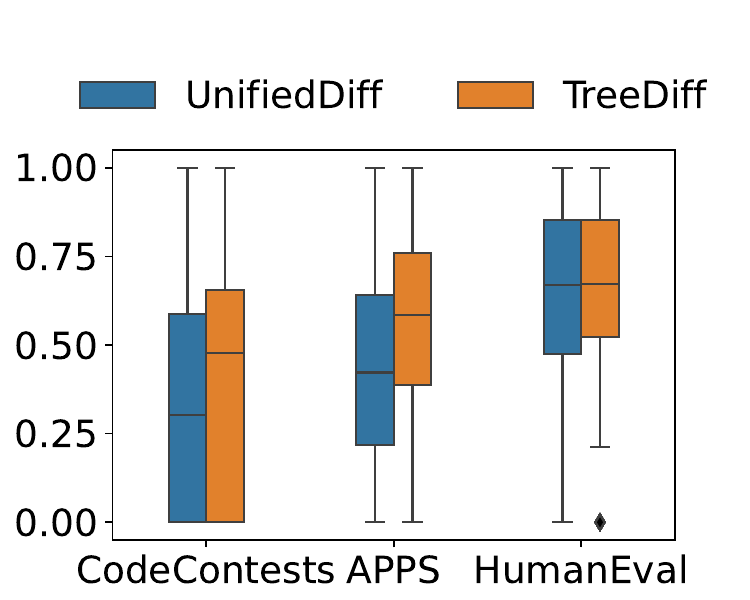}
    \caption{Mean}
  \end{subfigure}
  \hfill
  \begin{subfigure}{0.3\linewidth}
    \centering
    \includegraphics[width=\linewidth]{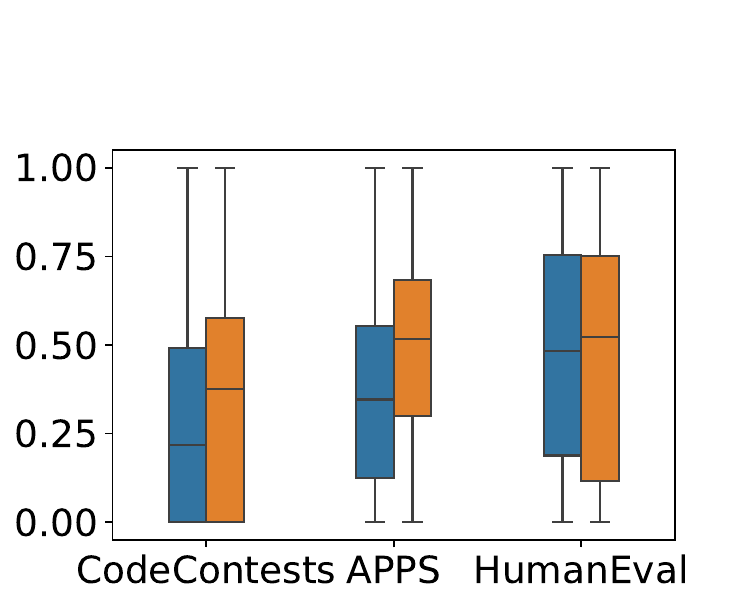}
    \caption{Mean Worst}
  \end{subfigure}
  \hfill
  \begin{subfigure}{0.3\linewidth}
    \centering
    \includegraphics[width=\linewidth]{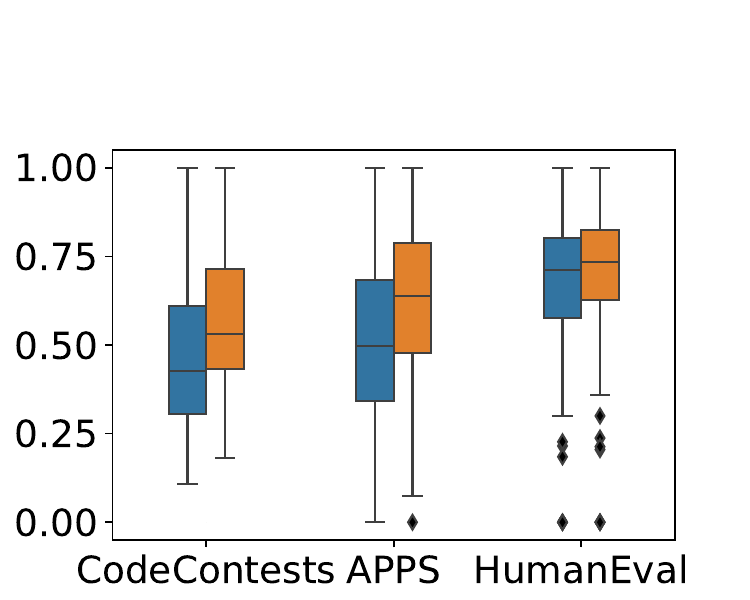}
    \caption{Pair Mean}
  \end{subfigure}
  \vspace{0mm}
  \caption{RQ1.3: Structural Similarity (United\_Diff \& Tree\_Diff).}
  \label{fig: structural_similarity}
\end{figure}

Structural similarity measures the codes' similarity based on their AST.
In our experiment, the structural similarity is mainly measured by the tool \texttt{pycode\_similar} with two different settings, namely United\_Diff and Tree\_Diff (more details in Section~\ref{subsec:similaritymeasurement}).
For the five code candidates for each coding problem, we use the first code candidate as a reference and calculate the structural similarity between the first candidate with the remaining four candidates under United\_Diff and Tree\_Diff settings.
We also calculate the structural similarity with code candidates in pairs, with a total of 20 pair mean values.
Thus, each problem has four mean values and 20 pair mean values for United\_Diff and Tree\_Diff respectively, with each value indicating a structural similarity measure.
We use the mean of these four values, the worst of them, and their pair mean values (i.e., the smallest value for United\_Diff and Tree\_Diff) to represent each problem's structural similarity.
Fig \ref{fig: structural_similarity} shows the distribution of United\_Diff and Tree\_Diff for all the problems in each dataset.
Table~\ref{tab:structuralsimilarity} shows the mean, mean worst values, and pair mean values under United\_Diff and Tree\_Diff settings for all the coding problems (the mean value inside each bar in the figures) in a dataset.


\begin{table}[h!]\scriptsize
\centering
\caption{RQ1.3: Structural similarity. 
}
\vspace{0mm}
\begin{tabular}{l l r r r }
\toprule
Structural Similarity & Metric & CodeContests  & APPS & HumanEval \\
\midrule
\multirow{2}{*}{United\_Diff} & Mean value & 0.33 & 0.43 & 0.60 \\
& Mean worst value & 0.27 & 0.35 & 0.47\\
&  Pair mean value & 0.46 & 0.52 & 0.67\\
\midrule
\multirow{2}{*}{Tree\_Diff} & Mean value & 0.41 & 0.54 & 0.62\\
&  Mean worst value & 0.33 & 0.47 & 0.48\\
&  Pair mean value & 0.56 & 0.63 & 0.70\\
\bottomrule
\end{tabular}
\label{tab:structuralsimilarity}
\end{table}



We observe that the code candidates generated from the same instruction show great similarity in structure.
Specifically,
the mean values are 0.33, 0.43, and 0.60 under the United\_Diff setting, and 0.41, 0.54, and 0.62 under Tree\_Diff setting for CodeContests, APPS, and HumanEval, respectively.

For the three datasets, 
we could see from Table~\ref{tab:structuralsimilarity} that the lowest values under United\_Diff and Tree\_Diff happen for the CodeContests dataset. 
By contrast, the largest values under the two settings both happen for HumanEval.
This indicates that ChatGPT is most unstable in structure for the code generation tasks in CodeContests, and most stable for HumanEval. 
We further explore the correlation between different similarities and task features in RQ4.

\begin{tcolorbox}
\textbf{\underline{Answer to RQ1.3:}} Code candidates show high structural similarity under UnitedDiff and TreeDiff settings.
We observe that the code candidates generated from the same instruction have high similarity in structure. 
Specifically,
the mean values are 0.33, 0.43, and 0.60 under the United\_Diff setting, and 0.41, 0.54, and 0.62 under Tree\_Diff setting for CodeContests, APPS, and HumanEval, respectively.

\end{tcolorbox}

\subsection{RQ2: Influence of Temperature}
The default temperature of ChatGPT is 1\footnote{\url{https://platform.openai.com/docs/api-reference/chat/create\#chat-create-temperature}}.
This RQ explores whether the code generation non-determinism of ChatGPT changes with the temperature changes.
We use identical measurements as in RQ1.
We show our experiment results on CodeContests only.
Results for other datasets are on our homepage~\cite{homepage}.

\begin{table}[h!]\scriptsize
\caption{RQ2: Influence of temperature (CodeContests).}
\vspace{0mm}
    \begin{tabular}{l |r r r r r r}
        \toprule
        \multirow{2}{*}{Temperature} & \multicolumn{6}{c}{Test Pass Rate} \\
        \cmidrule{2-7}

        & Mean value& Mean variance & Mean max diff& Max  diff& Ratio of worst cases& \\
        \midrule
        0 & 0.15 & 0.01 & 0.11 & 1.00 & 1.82\%\\
        0.5 & 0.16 & 0.02 & 0.15 & 1.00 & 2.42\%\\
        1 & 0.16 & 0.03 & 0.24 & 1.00 & 3.64\%\\

        \midrule
        \multirow{2}{*}{Temperature} & \multicolumn{3}{c}{OER}& \multicolumn{3}{c}{OER (no ex.)} \\
        \cmidrule{2-7}
         & Mean value &  Ratio of worst cases & Pair mean value & Mean value & Ratio of worst cases & Pair mean value \\

        \midrule
        0 & 0.37 & 43.64\% & 0.59 & 0.27 & 54.55\% & 0.46\\
        0.5 & 0.18 & 62.42\% & 0.37 & 0.13 & 68.48\% & 0.28\\
        1 & 0.09 & 75.76\% & 0.27 & 0.06  & 81.21\% & 0.19\\

        \midrule
        \multirow{2}{*}{Temperature} & \multicolumn{3}{c}{LCS}& \multicolumn{3}{c}{LED} \\
        \cmidrule{2-7}
         & Mean value & Mean worst value & Pair mean value & Mean value & Mean worst value & Pair mean value\\

        \midrule
        0 & 0.61 & 0.44 & 0.62 & 23.45 & 35.87 & 22.31\\
        0.5 & 0.33 & 0.23 & 0.34 & 44.48 & 62.02 & 44.89\\
        1 & 0.22 & 0.16 & 0.23 & 58.80 & 77.46 & 58.86\\

        \midrule
        \multirow{2}{*}{Temperature} & \multicolumn{3}{c}{United\_Diff} & \multicolumn{3}{c}{Tree\_Diff} \\
        \cmidrule{2-7}
         & Mean value & Mean worst value & Pair mean value & Mean value & Mean worst value & Pair mean value\\

        \midrule
        0 & 0.41 & 0.39 & 0.67 & 0.50 & 0.46 & 0.74\\
        0.5 & 0.61 & 0.49 & 0.63 & 0.69 & 0.58 & 0.71\\
        1 & 0.33 & 0.27 & 0.46 & 0.41 & 0.33 & 0.56\\

        \bottomrule
    \end{tabular}
\label{table: temperature(CodeContests)}
\end{table}

Table~\ref{table: temperature(CodeContests)} shows the results.
Overall, we observe that when \textit{temperature}=0, ChatGPT has better determinism than the default configuration (\textit{temperature}=1) for all three types of similarities.
However, setting the temperature to 0 does not completely avoid non-determinism.
Take OER as an example,
there are still 43.64\% (CodeContests), 27.40\% (APPS), and 18.29\% (HumanEval) of problems with no equal test output among the five code candidates.
This is contrary to many people's belief that setting the temperature to 0 can make ChatGPT deterministic~\cite{deng2023large,bhavya2022analogy,liang2023code}, because 
when setting the temperature to 0, the model applies greedy sampling which should indicate full determinism, with the logit value for the next token being a pure function of the input sequence and the model weights.
The reason for such non-determinism with the temperature being zero is still controversial~\cite{gptrandomness},
with different hypotheses such as floating point, unreliable GPU calculations, 
and its sparse MoE architecture failing to enforce per-sequence determinism \cite{puigcerver2023sparse, la2023arrt}.
The details for all the non-deterministic coding tasks and their test outputs with temperature=0 are on our homepage~\cite{homepage}.


When \textit{temperature}=0.5, we observe that
ChatGPT tends to generate code candidates that are more deterministic than temperature=1, but less deterministic than temperature=0.
This is as expected because the higher temperature brings more creativity to ChatGPT and affects its ability to generate similar code (as can be observed from the other measurements, such as LCS and LED).
Nevertheless, we observe that the value of test pass rates among the three different temperatures are similar, which indicates that low temperature might be a better choice given the comparable test pass rate and the low degree of non-determinism. 

\begin{tcolorbox}
\textbf{\underline{Answer to RQ2:}}
Contrary to the widely held belief (and common practices),
setting the temperature to 0 does not guarantee determinism in code generation, although it indeed brings 
more determinism than the default configuration (\textit{temperature}=1) for all three types of similarities.
We also observe that the values of test pass rate among the three different temperatures are similar, indicating that low temperature might be a better choice for code generation tasks.
\end{tcolorbox}

\subsection{RQ3: Non-determinism Comparison with Top Candidates in the Same Prediction}

RQ1 and RQ2 compare the similarity of 5 code candidates generated in multiple requests.
Each candidate is the top candidate in each request.
However, ChatGPT can also generate 5 code candidates within the same request (the top 5 candidates ranked by their predictive probabilities). 
This RQ compares the non-determinism degree of code candidates for the two request configurations mentioned above (with temperature = 1 and temperature =0).
Table~\ref{table: request way(CodeContests)} shows the results for CodeContests, the results for the two other datasets are on our homepage~\cite{homepage}.
For ease of presentation, 
we use R1 to refer to one-time requests, and R2 to refer to multiple requests.

Our results reveal that 
when setting temperature=1, it is difficult to tell which way of requesting is more deterministic.
For semantic similarity, R1 and R2's performance are similar among three datasets. 
Code candidates requested in R1 are slightly more random than those requested in R2 in terms of syntactic similarity since those requested in R1 have lower LCS values and higher LED values.
However, code candidates requested in R1 are slightly more stable than those requested in R2 when it comes to similarity because those requested in R1 have higher structural similarity values in both United\_Diff and Tree\_Diff settings.

When temperature is 0, the difference between the two request ways is obvious.
Code Candidates requested by R1 show higher determinism than those requested by R2.
When requesting by R1, the ratio of worst cases, where max diff is close to 0 (1.20\%), and the OER and OER (no ex.) are higher than R2 and close to 1.
The LCS values are higher than the values under other temperatures and LED values are lower than the values under other temperatures, which indicates higher determinism.
Among the three datasets, the structural similarity values are also higher than the values in other temperatures, which means the code candidates are more close to each other in terms of their AST structure.

\begin{table}[h!]\scriptsize
\caption{RQ3: Similarity for different request ways (CodeContests), where t represents the temperature setting.}
\vspace{0mm}
    \begin{tabular}{l |r r r r r r}
        \toprule
        \multirow{2}{*}{Request} & \multicolumn{6}{c}{Test Pass Rate} \\
        \cmidrule{2-7}

        Way & Mean value& Mean variance & Mean max diff& Max  diff & Ratio of worst cases& \\
        \midrule
        R1 (t=1) & 0.17 & 0.03 & 0.28 & 1.00 & 8.70\% \\
        R2 (t=1) & 0.16 & 0.03 & 0.24 & 1.00 & 3.64\% \\
        R1 (t=0) & 0.18 & 0.00 & 0.00 & 0.00 & 1.20\%\\
        R2 (t=0) & 0.15 & 0.01 & 0.11 & 1.00 & 1.82\%\\

        \midrule
        \multirow{2}{*}{Request} & \multicolumn{3}{c}{OER}& \multicolumn{3}{c}{OER (no ex.)} \\
        \cmidrule{2-7}
         & Mean value &  Ratio of worst cases & Pair mean value & Mean value & Ratio of worst cases & Pair mean value \\

        \midrule
        R1 (t=1) & 0.09 & 76.09\% & 0.27 & 0.04 & 83.70\% & 0.18 \\
        R2 (t=1) & 0.09 & 75.76\% & 0.27 & 0.06 & 81.21\% & 0.19 \\
        R1 (t=0) & 1.00 & 1.20\% & 1.00 & 0.81 & 12.05\% & 0.81\\
        R2 (t=0) & 0.37 & 43.64\% & 0.59 & 0.27 & 54.55\% & 0.46\\

        \midrule
        \multirow{2}{*}{Request} & \multicolumn{3}{c}{LCS}& \multicolumn{3}{c}{LED} \\
        \cmidrule{2-7}
         Way & Mean value & Mean worst value & Pair mean value & Mean value & Mean worst value & Pair mean value\\

        \midrule
        R1 (t=1) & 0.21 & 0.15 & 0.20 & 61.30 & 82.73 & 63.09\\
        R2 (t=1) & 0.22 & 0.16 & 0.23 & 58.80 & 77.46 & 58.86\\
        R1 (t=0) & 1.00 & 1.00 & 1.00 & 0.00 & 0.00 & 0.00\\
        R2 (t=0) & 0.61 & 0.44 & 0.62 & 23.45 & 35.87 & 22.31\\

        \midrule
        \multirow{2}{*}{Request} & \multicolumn{3}{c}{United\_Diff} & \multicolumn{3}{c}{Tree\_Diff} \\
        \cmidrule{2-7}
         Way & Mean value & Mean worst value & Pair mean value & Mean value & Mean worst value & Pair mean value\\

        \midrule
        R1 (t=1) & 0.98 & 0.98 & 0.98 & 0.98 & 0.98 & 0.98\\
        R2 (t=1) & 0.33 & 0.27 & 0.46 & 0.41 & 0.33 & 0.56\\
        R1 (t=0) & 1.00 & 1.00 & 1.00 & 1.00 & 1.00 & 1.00\\
        R2 (t=0) & 0.41 & 0.39 & 0.67 & 0.50 & 0.46 & 0.74\\
        
        \bottomrule
    \end{tabular}
\label{table: request way(CodeContests)}
\end{table}

\begin{tcolorbox}
\textbf{\underline{Answer to RQ3:}}
Under default temperature, the top-5 code candidates from one single request have similar non-determinism with the 5 top-1 candidates from different requests for ChatGPT when the temperature is 1 (default temperature of ChatGPT), but higher determinism when 
the temperature is 0.
\end{tcolorbox}

\subsection{RQ4: Coding Tasks Features and Non-determinism Degree}
\label{subsec:correlationbetweencodetasks}

Our previous experiments demonstrate that there are many non-determinisms in ChatGPT in code generation.
This RQ investigates what affects such non-determinism by checking the correlation between characteristics of coding tasks and similarity metric values.
We use three datasets for this RQ.
For all the datasets, we consider \textit{description length} as one of their extrinsic features.
Because only the CodeContests dataset has various extrinsic features for each coding task, including \textit{difficulty}, \textit{time limit}, and \textit{CF rating}, we consider these features as extrinsic features for the CodeContest dataset as well.
Although APPS does have \textit{difficulty} features, the \textit{difficulty} features in APPS are shown as categories, namely, `introductory', `interview', and `competition', which makes it hard to map them into numerical values.
Therefore, our experiment does not include \textit{difficulty} as an extrinsic feature for the APPS dataset.

In CodeContests, the \textit{CF rating} of a problem is a quantitative measure that represents the problem's relative difficulty level compared to other problems on the Codeforces platform.
The \textit{difficulty} of a problem is a qualitative measure that indicates the problem's level of complexity and the programming knowledge and skills required to solve it. 
The \textit{timeout} indicates the program's maximum running time limitation. 
In addition, we also consider description length (i.e., number of characters) for each coding task.
Note that in this section, we only focus on correlation analysis, and we do not aim to obtain any causal conclusions.

\begin{figure}[t]
\hspace{-4mm}\includegraphics[width=1.0\linewidth]{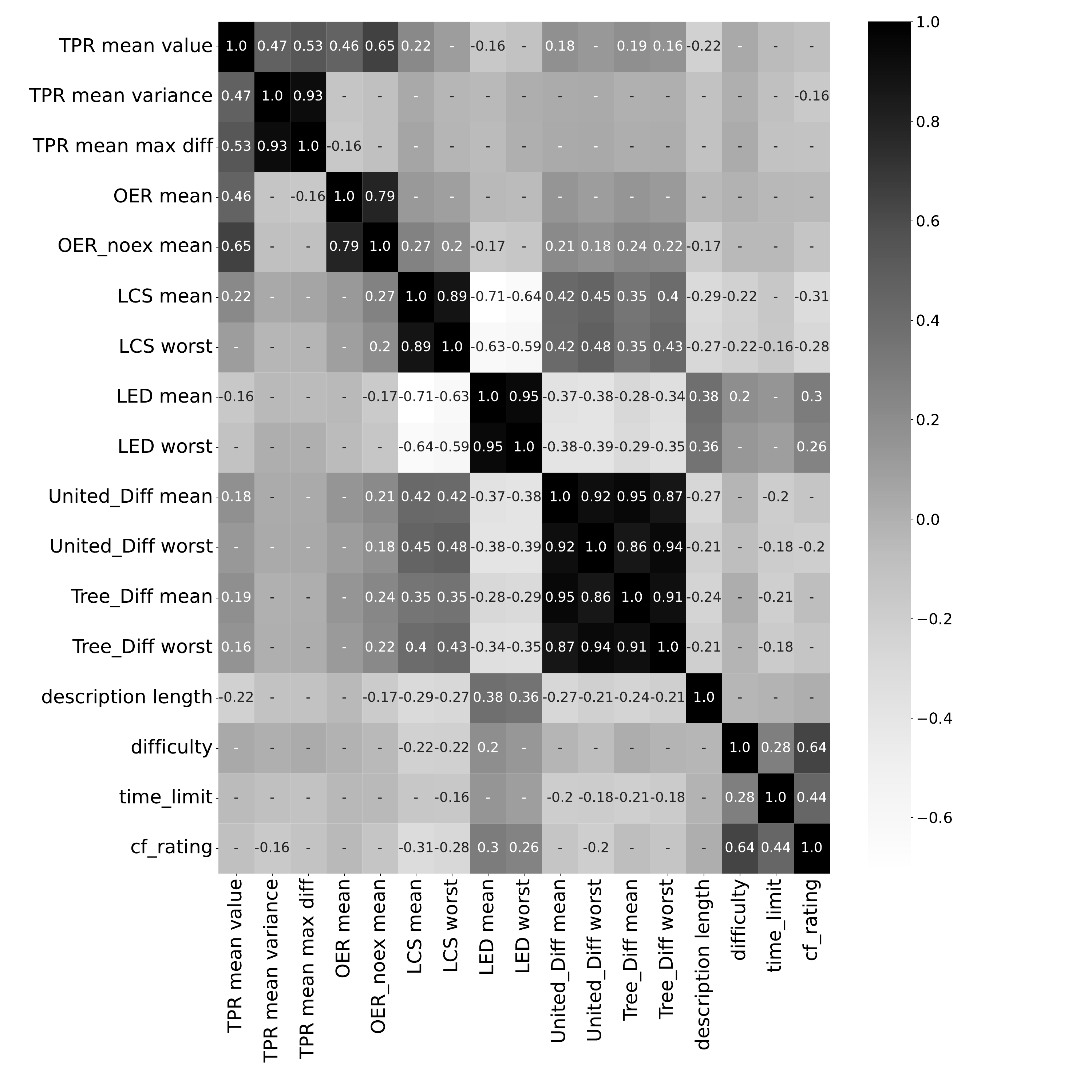}
\vspace{-5mm}
\caption{RQ4: Correlations between coding tasks and non-determinism (CodeContests, temperature=1). Only significant correlations will be displayed on the heatmap, while the insignificant correlations (i.e. p-value \(>\) 0.05) are masked by `-'.
}
  \label{fig: heatmap_CodeContests_1}
\end{figure}

Figure~\ref{fig: heatmap_CodeContests_1} shows the results for code problems in CodeContests under temperature=1.
The rest figures can be found on our homepage~\cite{homepage}.
We observe that 
description length has a negative correlation with most of the measurements, except LED.
This means that problems with longer descriptions tend to generate code with more randomness.
We suspect that this is because a longer description may reduce ChatGPT's understanding of the coding requirements.
With longer descriptions, different code candidates tend to be uniformly worse in their pass rates. 
Moreover, the description length has a negative correlation with LCS and structural measurements and a positive correlation with LED, which means that problems with longer descriptions tend to yield more inconsistent code candidates in syntax and structure.
For temperature = 0, we observe that 
description length still has a negative correlation with most of the measurements, except LED, which is similar to the correlation result under temperature=1.

The \textit{difficulty} has a positive correlation with the LED and a negative correlation with LCS, which means that the problem with a higher \textit{difficulty} level has high non-determinism in syntax.
Similar to \textit{difficulty}, \textit{CF rating} also has a positive correlation with the LED and a negative correlation with LCS.

In the following, we provide some specific examples to further illustrate our observations above. 
In exploring the relationship between the length of a code problem description and the degree of non-determinism, two contrasting examples in the CodeContests dataset corroborate our findings.
The first example, `1599\_E. Two Arrays', with a description length of 2149, show a pattern that code generation with a longer description code problem has a higher degree of non-determinism.
Below is the description of the first code problem, where we present only the core part of the description due to the extensive length of the overall content.


\begin{tcolorbox}[colback=gray!20, 
                 colframe=white, 
                 boxrule=0pt, 
                 arc=0pt, 
                 top=3pt, bottom=3pt, left=5pt, right=5pt]
\textbf{1599\_E. Two Arrays}

You are given two integer arrays of length N, A1, and A2. You are also given Q queries of 4 types: 

$1\ k\ l\ r\ x: set Ak_i:=min(Ak_i, x)$ for each $l \leq i \leq r$.

$2\ k\ l\ r\ x: set Ak_i:=max(Ak_i, x)$ for each $l \leq i \leq r$.

$3\ k\ l\ r\ x: set Ak_i:=Ak_i+x$ for each $l \leq i \leq r$.

$4\ l\ r: find the (\sum_{i=l}^r F(A1_i+A2_i)) \% (10^9+7)$ 

where $F(k)$ is the k-th Fibonacci number $(F(0)=0, F(1)=1, F(k)=F(k-1)+F(k-2))$, and $x \% y$ denotes the remainder of the division of $x$ by $y$.
You should process these queries and answer each query of the fourth type.
\end{tcolorbox}

This problem exhibits high non-determinism, as indicated by its measurement results across multiple tests (i.e., the test case rate variance is 0.13, the OER value is zero, the LCS mean value is 0.15, the mean LED value is 111.5, and both the United\_Diff and Tree\_Diff values are zero), suggesting a rather high fluctuation.
The detailed description potentially covers a wide array of scenarios, which may distract the attention from LLMs, which results in inconsistent test results and higher non-determinism.

The second example `1575\_M. Managing Telephone Poles', with a description length of 1511, shows a pattern that a shorter description leads to more stability in code generation.
Below is the description of the second code problem, where we present only the core part of the description due to the extensive length of the overall content.

\begin{tcolorbox}[colback=gray!20, 
                 colframe=white, 
                 boxrule=0pt, 
                 arc=0pt, 
                 top=3pt, bottom=3pt, left=5pt, right=5pt]
\textbf{1575\_M. Managing Telephone Poles}

Mr. Chanek's city can be represented as a plane. He wants to build a housing complex in the city.
There are some telephone poles on the plane, which is represented by a grid of size $(n + 1) × (m + 1)$. 

There is a telephone pole at $(x, y)$ if $a_{x, y} = 1$.
For each point $(x, y)$, define $S(x, y)$ as the square of the Euclidean distance between the nearest pole and $(x, y)$. 

Formally, the square of the Euclidean distance between two points $(x_1, y_1)$ and $(x_2, y_2)$ is $(x_2 - x_1)^2 + (y_2 - y_1)^2$.

To optimize the building plan, the project supervisor asks you the sum of all $S(x, y)$ for each $0 \leq x \leq n$ and $0 \leq y \leq m$. Help him by finding the value of $\sum_{x=0}^{n} {\sum_{y=0}^{m} {S(x, y)}}$.
\end{tcolorbox}

The test pass rates are consistently 1.0 across all tests, with a variance of 0.0, showing no deviation in the generated code candidates.
The LCS mean value is 0.74, and the LED mean value is 3.5, which indicates a high syntactical stability.
Structural similarity is 0.21 and 0.38 under United\_Diff and Tree\_Diff settings, which shows the code candidates still vary in their AST.
Here, the shorter description does not introduce ambiguity but rather lets ChatGPT focus on critical details, leading to a uniform understanding of the code problem and better generation performance.


\begin{tcolorbox}
\textbf{\underline{Answer to RQ4:}} A coding task with a longer description and higher difficulty tends to suffer from more non-determinism in the generated code in terms of code syntax and structure.
The generated code also tends to be more buggy.
\end{tcolorbox}

\subsection{RQ5: GPT-4 vs. GPT-3.5}

GPT-4 is believed to be ``more reliable, creative, and able to handle much more nuanced instructions than GPT-3.5''~\cite{openai2023gpt4}.
This research question compares GPT-3.5 and GPT-4 in
the non-determinism degree of code generation.
To answer this research question, we keep the
default setting and use all the measurements listed in RQ1.
In this paper we report the results only on the CodeContests dataset (with temperature=1).
For the results in the other two datasets, we list them on our homepage~\cite{homepage}.

For temperature=1, we can observe that GPT-4 is slightly more deterministic than GPT-3.5, with lower test pass rate variance, lower ratio of worst cases, lower OER and OER (no ex.), lower LCS, higher LED, and lower structural similarity under two settings.
However, for temperature=0, the analysis, as evidenced by the results in tables comparing GPT-4 across CodeContests, APPS, and HumanEval datasets, demonstrates that GPT-4's non-determinism is pronounced and largely parallels that of GPT-3.5.
Across these datasets, similarity metrics indicate comparable levels of non-determinism across three different evaluation methods.

\begin{table}[h!]\scriptsize
\caption{RQ5: Non-determinism of GPT-4 v.s. GPT-3.5 (CodeContests)}
\vspace{0mm}
    \begin{tabular}{l |r r r r r r}
        \toprule
        \multirow{2}{*}{Model} & \multicolumn{6}{c}{Test Pass Rate} \\
        \cmidrule{2-7}

         & Mean value& Mean variance & Mean max diff& Max  diff& Ratio of worst cases& \\
        \midrule
        GPT-4 (t=1) & 0.14 & 0.01 & 0.09 & 1.00 & 1.21\% \\
        GPT-3.5 (t=1) & 0.16 & 0.03 & 0.24 & 1.00 & 3.64\%\\
        GPT-4 (t=0) & 0.14 & 0.01 & 0.08 & 1.00 & 1.21\% \\
        GPT-3.5 (t=0) & 0.15 & 0.01 & 0.11 & 1.00 & 1.82\%\\

        \midrule
        \multirow{2}{*}{Model} & \multicolumn{3}{c}{OER}& \multicolumn{3}{c}{OER (no ex.)} \\
        \cmidrule{2-7}
         & Mean value &  Ratio of worst cases & Pair mean value & Mean value & Ratio of worst cases & Pair mean value \\

        \midrule
        GPT-4 (t=1) & 0.35 & 46.06\% & 0.58 & 0.25 & 55.76\% & 0.46 \\
        GPT-3.5 (t=1) & 0.09 & 75.76\% & 0.27 & 0.06 & 81.21\% & 0.19\\
        GPT-4 (t=0) & 0.37 & 41.21\% & 0.59 & 0.27 & 52.73\% & 0.46\\
        GPT-3.5 (t=0) & 0.37 & 43.64\% & 0.59 & 0.27 & 54.55\% & 0.46\\
                
        \midrule
        \multirow{2}{*}{Model} & \multicolumn{3}{c}{LCS}& \multicolumn{3}{c}{LED} \\
        \cmidrule{2-7}
         & Mean value & Mean worst value & Pair mean value & Mean value & Mean worst value & Pair mean value\\

        \midrule
        GPT-4 (t=1) & 0.61 & 0.45 & 0.62 & 24.54 & 39.74 & 24.81 \\
        GPT-3.5 (t=1) & 0.22 & 0.16 & 0.23 & 58.80 & 77.46 & 58.86 \\
        GPT-4 (t=0) & 0.61 & 0.44 & 0.61 & 24.45 & 40.14 & 24.12 \\
        GPT-3.5 (t=0) & 0.61 & 0.44 & 0.62 & 23.45 & 35.87 & 22.31\\

        \midrule
        \multirow{2}{*}{Model} & \multicolumn{3}{c}{United\_Diff} & \multicolumn{3}{c}{Tree\_Diff} \\
        \cmidrule{2-7}
         & Mean value & Mean worst value & Pair mean value & Mean value & Mean worst value & Pair mean value\\

        \midrule
        GPT-4 (t=1) & 0.78 & 0.68 & 0.79 & 0.82 & 0.74 & 0.84 \\
        GPT-3.5 (t=1) & 0.33 & 0.27 & 0.46 & 0.41 & 0.33 & 0.56 \\
        GPT-4 (t=0) & 0.78 & 0.68 & 0.79 & 0.83 & 0.75 & 0.84\\
        GPT-3.5 (t=0) & 0.41 & 0.39 & 0.67 & 0.50 & 0.46 & 0.74\\

        \bottomrule
    \end{tabular}
\label{table: gpt-4 comparison(CodeContests)}
\end{table}

\begin{tcolorbox}
\textbf{\underline{Answer to RQ5:}}
The non-determinism issue of GPT-4 is lightly less severe than GPT-3.5 under temperature=1, while the non-determinism issue of GPT-4 is similar to GPT-3.5 under temperature=0.
\end{tcolorbox}

\subsection{RQ6: Influence of Prompt Engineering Strategies on the Non-determinism}

This research question explores how different prompt engineering strategies influence the degree of non-determinism in code generation. 
We design two extra prompts in addition to the default one used for previous RQs. 
The first prompt is ``Generate Python3 code (Markdown), make the code as concise as possible''. This prompt aims to lead ChatGPT to 
generate short and concise programs, which may make the results more deterministic.
The second prompt is ``Generate Chain-of-Thought steps of how to solve the problem first, and then generate Python3 code (Markdown)'', thereby demanding an initial conceptual explanation followed by the code.
Then, each prompt is followed by the code problem description.
In the following, 
we use `Concise prompt' to refer to the first prompt engineering strategy, and use `CoT prompt' to refer to the second one for short.

\begin{table}[h!]\scriptsize
\caption{RQ6: Prompt engineering techniques (CodeContests), where t refers to temperature.}
\vspace{0mm}
    \begin{tabular}{l |r r r r r r}
        \toprule
        \multirow{2}{*}{Prompt} & \multicolumn{6}{c}{Test Pass Rate} \\
        \cmidrule{2-7}

         & Mean value& Mean variance & Mean max diff& Max  diff& Ratio of worst cases& \\
        \midrule
        Concise (t=1) & 0.15 & 0.02 & 0.19 & 1.00 & 3.64\% \\
        Base (t=1) & 0.16 & 0.03 & 0.24 & 1.00 & 3.64\% \\
        CoT (t=1) & 0.15 & 0.02 & 0.19 & 1.00 & 3.64\% \\
        Concise (t=0) & 0.16 & 0.01 & 0.10 & 1.00 & 0.61\%\\
        Base (t=0) & 0.15 & 0.01 & 0.11 & 1.00 & 1.82\% \\
        CoT (t=0) & 0.19 & 0.02 & 0.15 & 1.00 & 1.82\%\\

        \midrule
        \multirow{2}{*}{Prompt} & \multicolumn{3}{c}{OER}& \multicolumn{3}{c}{OER (no ex.)} \\
        \cmidrule{2-7}
         & Mean value &  Ratio of worst cases & Pair mean value & Mean value & Ratio of worst cases & Pair mean value \\

        \midrule
        Concise (t=1) & 0.10 & 76.36\% & 0.26 & 0.06 & 81.82\% & 0.17 \\
        Base (t=1) & 0.09 & 75.76\% & 0.27 & 0.06 & 81.21\% & 0.19 \\
        CoT (t=1) & 0.10 & 73.94\% & 0.26 & 0.08 & 80.0\% & 0.19\\
        Concise (t=0) &  0.39 & 41.82\% & 0.63 & 0.31 & 49.09\% & 0.54\\
        Base (t=0) & 0.37 & 43.64\% & 0.59 & 0.27 & 54.55\% & 0.46 \\
        CoT (t=0) & 0.28 & 46.06\% & 0.50 & 0.19 & 54.55\% & 0.36\\
        
        \midrule
        \multirow{2}{*}{Prompt} & \multicolumn{3}{c}{LCS}& \multicolumn{3}{c}{LED} \\
        \cmidrule{2-7}
         & Mean value & Mean worst value & Pair mean value & Mean value & Mean worst value & Pair mean value\\

        \midrule
        Concise (t=1) & 0.22 & 0.16 & 0.22 & 61.53 & 83.01 & 62.52 \\
        Base (t=1) & 0.22 & 0.16 & 0.23 & 58.80 & 77.46 & 58.86\\
        CoT (t=1) & 0.23 & 0.15 & 0.23 & 59.55 & 77.68 & 57.05 \\
        Concise (t=0) & 0.70 & 0.53 & 0.71 & 11.77 & 20.55 & 12.14 \\
        Base (t=0) & 0.61 & 0.44 & 0.62 & 23.45 & 35.87 & 22.31 \\
        CoT (t=0) & 0.38 & 0.24 & 0.39 & 39.31 & 58.28 & 39.81\\
        
        \midrule
        \multirow{2}{*}{Prompt} & \multicolumn{3}{c}{United\_Diff} & \multicolumn{3}{c}{Tree\_Diff} \\
        \cmidrule{2-7}
         & Mean value & Mean worst value & Pair mean value & Mean value & Mean worst value & Pair mean value\\

        \midrule
        Concise (t=1) & 0.44 & 0.34 & 0.48 & 0.54 & 0.42 & 0.59 \\
        Base (t=1) &  0.33 & 0.27 & 0.46 & 0.41 & 0.33 & 0.56 \\
        CoT (t=1) & 0.45 & 0.35 & 0.51 & 0.55 & 0.43 & 0.61 \\
        Concise (t=0) & 0.83 & 0.74 & 0.84 & 0.88 & 0.82 & 0.89 \\
        Base (t=0) & 0.41 & 0.39 & 0.67 & 0.50 & 0.46 & 0.74 \\
        CoT (t=0) & 0.71 & 0.58 & 0.72 & 0.78 & 0.67 & 0.79\\
        
        \bottomrule
    \end{tabular}
\label{table: complexity(CodeContests)}
\end{table}

The results in Table~\ref{table: complexity(CodeContests)} show that 
for temperature=1, the difference of non-determinism between different prompt engineering techniques is not very obvious in the three datasets. 
With more instruction information provided in the prompt, Concise and CoT prompts have similar performance with each other.
However, under temperature=0,
in CodeContests, requests with CoT prompt show high mean test pass rates but this kind of prompt suffers from high randomness.
Compared with the Base prompt and Concise prompt, the CoT prompt has a higher mean-variance (0.02), higher mean maximum difference (0.15), and a rather higher ratio of worst cases (1.82\%).
Also, the results in OER and OER (no ex.) show that CoT's mean value of OER and OER (no ex.) are lower than Base and Concise, which can also be told from the high ratio of worst cases in both OER and OER (no ex.) with 46.06\% and 54.55\%.
Opposite from CoT, code candidates generated from Concise prompt are more semantically deterministic.
Code candidates generated by the CoT prompt have a low mean LCS value (0.38) and high LED value (39.31), while those generated from the Concise prompt have a high mean LCS value (0.07) and low LED value (11.77).
The other measurements in LCS and LED also support the above phenomenon. 
When it comes to structural similarity, under two different measurement settings, code candidates generated from the CoT prompt have significantly higher randomness than the code generated from Concise prompt.
Our experiment results show a similar situation in both APPS and HumanEval, where code generated from the Concise prompt ends up way more deterministic than code generated from the CoT prompt.

\begin{tcolorbox}
\textbf{\underline{Answer to RQ6:}}
Under temperature=1, the difference in non-determinism among different prompt engineering techniques is not obvious.
When setting temperature=0, the code candidates generated from the Concise prompt are more deterministic than our Base prompt, while those code candidates generated from the CoT prompt suffer from higher randomness than our Base prompt.
\end{tcolorbox}

\section{Threats to Validity}

The threats to \textit{internal} validity mainly lie in the implementation of our experiment and result analysis.
To reduce the first threat, we checked our code twice, once during the experiment stage, and once during the record analysis stage.
To reduce the second threat, the two authors independently analyzed the experiment results and drew experimental conclusions separately.
Once their analysis results were different, the third author discussed with them to determine the final result.

The threats to \textit{external} validity mainly lie in the datasets, GPT versions, and prompt design in our study.
To reduce the threat in datasets, we use three diverse datasets that are widely used in code generation tasks.
Additionally, the problems in our dataset are from different contests with different difficulties.
For example, CodeContests is the most challenging dataset, while HumanEval is the easiest, in terms of the average difficulty of coding problems. 
To reduce the threat in GPT versions, we consider the two newest versions of GPT: GPT-3.5 and GPT-4, and compare their non-determinism from multiple aspects. 
To reduce the threat of prompt design, we use the most typical prompts that are the most widely used in LLM-based code generation and design an RQ to study their influence on non-determinism.

Another primary concern highlighted in our analysis revolves around the operationalization of semantic, syntactic, and structural similarities into measurable metrics for assessing code similarity.
The approach of measuring semantic similarity through the comparison of test execution outputs, while practical, presents a notable limitation.
It potentially oversimplifies the multifaceted nature of semantic similarity, which should ideally encapsulate the code's meaning and functionality rather than merely its output.
This method risks ignoring the intricate logic and diverse correct solutions that different pieces of code may offer.
To reduce the threat in measurement tools, we consider three types of similarities and choose at least two measurements for each type of similarity, and we also apply statistical analysis techniques to enhance our experiment results.
For the HumanEval dataset, we evaluate our measurement on an external testset, EvalPlus~\cite{liu2023your}.
The result shows that our measurements show similar evaluation results, which supports the robustness of our chosen measurements.



However, it is important to acknowledge certain \textit{limitations} within our study that may affect the breadth of its applicability and the generalizability of its findings.
Firstly, our analysis does not extend to the impact that different programming languages might have on the non-determinism of code generation.
Programming languages vary widely in syntax, semantics, and complexity, which can influence how LLMs like ChatGPT interpret and generate code, potentially affecting the degree of non-determinism in the output.
Secondly, our work only adopts a few methods for measuring code similarity.
There is no unified standard for measuring code similarity. 
It is challenging to cover all the code similarity measurements. 
Other methods include embedding-based similarity measure methods, using pre-trained code language models, such as CodeBERT~\cite{feng2020codebert} and GraphCodeBERT~\cite{guo2020graphcodebert}.
Thirdly, the influence of the prompt on non-determinism is not fully considered.
The specificity, clarity, and technical depth of prompts provided to ChatGPT can significantly influence the model's output, suggesting that prompts could be a crucial factor in understanding non-determinism.
Fourthly, our study focuses exclusively on ChatGPT.
While ChatGPT is a prominent LLM used for code generation, it is not the only one.
The landscape of LLMs is diverse, with models trained on different datasets, architectures, and objectives.
Therefore, our findings may not apply to other LLMs used for similar purposes.



\section{Related Work}
\subsection{Code Generation}

Code generation generates programs that need to satisfy all the constraints defined by the underlying task.
Usually, the constraints are represented in various forms, e.g. input/output pairs, examples, problem descriptions, partial programs, and assertions. Relatively early work includes deductive synthesis approaches \cite{manna1971toward, green1981application} and inductive synthesis approaches \cite{shaw1975inferring, biermann1978inference, summers1977methodology, smith1975pygmalion}. The deductive synthesis approach operated under the assumption that a comprehensive and precise formal specification of the user's desired intention would be provided. However, in many instances, this turned out to be just as intricate and challenging as creating the actual program. While the inductive synthesis approach was based on inductive specifications such as input/output pairs and examples etc, such as works on Lisp programs \cite{shaw1975inferring, biermann1978inference, summers1977methodology}, Pygmalion \cite{smith1975pygmalion} and more recently FlashFill \cite{gulwani2011automating}. More information could be found in a survey \cite{gulwani2017program}, which covers notable work on the development of program synthesis approaches.

In recent years, more and more researchers apply neural networks in code generation. Yin and Neubig \cite{yin2017syntactic} combine the grammar rules with the decoder and propose a syntax-driven neural architecture to improve code generation performance. Instead of RNN, Sun et al. \cite{sun2019grammar} propose a grammar-based structural CNN to capture the long dependency in code. Bolin et al. \cite{wei2019code} propose a dual learning framework that jointly trains the code generation model and code summarization model together to achieve better performance in both tasks. 
Xu et al.\cite{xu2022ide} present a user study in-IDE code generation, demonstrating challenges such as time efficiency, correctness, and code quality, as well as the willingness to use code generation tools from developers.

\subsection{Language Model for Code generation}

The triumph of transformers in natural language modeling \cite{brown2020language} has stimulated considerable interest among researchers in applying transformer models for code generation. Existing research on code generation models can be classified into three categories: sequence-based techniques, tree-based methods, and pre-trained models. 

\textbf{Sequence-based techniques} take code as a sequence of tokens and employ language models to produce source code one token at a time based on input descriptions. Ling et al. \cite{ling2016latent} propose a generative model for code generation along with a character level softmax and multi-pointer network to address the problem of generating code from a mixed language and structured specification, and receiving success in trading card games (Magic the Gathering and Hearthstone). Hashimoto et al. \cite{hashimoto2018retrieve} train a retrieval model with a noisy encoder-decoder to enable similar code retrieving, and then use the similar code as an additional input to improve the performance of the generator.

\textbf{Tree-based methods} generate a parse tree of the code, e.g. Abstract Syntax Tree (AST), based on the input description, and then convert the parse tree into the corresponding code. Dong et al. \cite{dong2016language} encode natural language utterances into vectors and generate their corresponding logical forms as trees using the LSTM model. Yin et al. \cite{yin2018tranx} propose a semantic parser `Tranx', which generates the tree-construction action sequence with a transition-based neural model, and constructs the AST from the action sequence.

\textbf{Pre-trained models} are obtained from training on massive data of source code, which could be later fine-tuned on certain datasets for code generation purposes. Encoder pre-trained models, such as CodeBERT \cite{feng2020codebert}, usually are trained with two objectives, i.e., Masked Language Modeling and Replaced Token Detection. During the fine-tuning phase, the input should be fed in the same way as the pre-training phrase, so that semantic relevance could be measured. Decoder pre-trained models are designed to predict the next token based on a given input context. GPT-series \cite{radford2018improving} are typical Decoder pre-trained models, and based on GPT, there are many efforts on code generation. Based on GPT-2, Lu et al. \cite{lu2021codexglue} provide CodeGPT for code completion and text-to-code generation. After GPT-3 was developed, CodeX\footnote{\url{https://openai.com/blog/openai-codex}} and GitHub Compilot\footnote{\url{https://github.com/features/copilot}} was created and released their beta version for trial by academia and industry. Due to neither Codex nor GitHub Copilot being open-sourced, there are several attempts to reproduce their performance, like PYCODEGPT-CERT \cite{zan2022cert}, CodeParrot\footnote{\url{https://huggingface.co/codeparrot/codeparrot}}, and GPT-CC\footnote{\url{https://github.com/CodedotAl/gpt-code-clippy}}. Encoder-decoder pre-trained models are composed of an encoder and a decoder. AlphaCode \cite{li2022competition}, which is pre-trained through GitHub repositories with 715.1 GB of code, uses an encoder-decoder transformer architecture. It achieves on average a ranking in the top 54\% in competitions with more than 5,000 participants in simulated evaluations.

ChatGPT, a language model developed by the team of OpenAI, has the potential to play a role in code generation. As it is widely known, ChatGPT offers a chat window to enable interaction in a conversational way. In addition to its powerful capabilities for natural language processing tasks, ChatGPT inherits the code generation capabilities from Codex and can perform even better, so the OpenAI team has announced the deprecation of Codex series models in its official documents. There are several research works that mentioned its ability in code-related areas, including mathematical capability \cite{frieder2024mathematical}, bug-solving capability \cite{surameery2023use}, and software testing \cite{jalil2023chatgpt}. ChatGPT's `Regenerate response' function demonstrates the diversity of its output, but at the same time, it also raises concerns about the consistency of its output given the same input. Currently, people are amazed by its superficial performance in terms of code generation, however, there is still no research work focused on the threat of non-determinism. Therefore, we think it is necessary to make a comprehensive evaluation of ChatGPT's ability in code generation. More detailed information could be found on its official website's blog~\cite{openaichat}.

\subsection{Non-determinism Handling in the Literature}

The non-determinism issue has been studied in traditional Deep Learning-related research: Pham et al.~\cite{pham2020problems} measure the influence of nondeterminism-introducing (NI)-factors in Deep Learning, and study the awareness of this variance among researchers and practitioners. However, the severity of the non-determinism threat in LLM-based coding studies remains unclear.

To understand how well LLM-based code generation papers handle the threat of non-determinism,
we collect research articles from Google Scholar with the query ‘code generation’ AND ‘Large Language Model’ in the past 2 years (from January 2022 to July 2023).
During the search, we search the full text of the paper (excluding citations and appendixes) for keywords, such as non-determinism and its synonyms, the number of experimental repetitions, and the variance of experimental results.
After locating these keywords, we manually combine the context to confirm whether the sentence means to declare that non-determinism exists in their study.
If the statement exists in the experimental section of the paper and the authors consider non-determinism in their experiment setting and result report, we classify it as considering non-determinism in the experimental design and mentioning non-determinism in the paper;
otherwise, if non-determinism is mentioned elsewhere without any actions to mitigate non-determinism, such as in the discussion section, we classify it as only mentioning non-determinism, but not considering this factor in the experiment.
If the above keywords are not mentioned in the paper, we read the full text of the paper to ensure that there are no sentences mentioning non-determinism in the paper.
If relevant non-determinism statements were encountered, we classify the paper using the above classification method and update our keyword library.
After ensuring that our keyword database is up to date and that the two search results are consistent, we searched all the papers twice to obtain our literature review data.

There are 107 papers obtained from Google Scholar according to their relevance rankings.
In this survey, we mainly focus on articles with experiments and exclude those with posters and visions only, which yields a set of 76 papers.
After an in-depth reading of the experimental design and discussion in each paper, we find that only 35.5\% (27/76) out of the 76 papers mention non-determinism or related terms (e.g., stability, randomness, and variance) in their papers.
Among them,
21.1\% (16/76) papers consider non-determinism in their experimental evaluation, including fixed random seeds, multiple runs of experiments with different fixed random seeds, and report results with error bars or standard deviation. 
In addition, 14.5\% (11/76) of the papers do not consider non-determinism in their experiments, but discuss the threat of non-determinism in their paper.

\section{Discussion}

In this section, we discuss the implications, trade-offs of non-determinism, and future research directions for code generation with LLMs.

\subsection{Implications for Software Developers and Researchers}

\textbf{For developers}, it is essential to recognize the limitations of ChatGPT and the potential risks of using generated code in production. 
If developers prefer a more stable code, they can use a smaller temperature but should keep in mind that even the smallest temperature (i.e., \textit{temperature}=0) could not guarantee the determinism.
Moreover, our observation on the correlation between the length of prompts and code correctness/non-determinism suggests the importance of prompt engineering. 
Developers should thoroughly test the generated code before deploying it, and even consider incorporating more robust testing and validation processes to ensure the determinism and reliability of the generated code.


\textbf{For researchers}, the variance of the generated code raises questions about the quality and validity of the results obtained from assessing LLMs in code generation.
If the code generated from ChatGPT is unstable, it can lead to non-reproducible results and unreliable conclusions.
Therefore, researchers should carefully consider the limitations of ChatGPT when designing experiments and interpreting results. 
To reduce the randomness caused by the non-determinism issue, researchers can report the average results, variance, or distribution from multiple requests.
Also, it is important to use different datasets, since our study finds that both the correctness and non-determinism of the generated code vary significantly from dataset to dataset. 
In addition, using a prompt with detailed instructions, a clear structure, and concrete response requirements would help to reduce randomness in generated code.


\subsection{Trade-off of non-determinism}


Our empirical study highlights the issue of non-determinism in code generation tasks when using ChatGPT. 
While we underscore the challenges this non-determinism introduces, particularly in terms of ensuring consistency and reliability in generated outputs, it is essential to also acknowledge the potential benefits that non-determinism brings, especially in the realm of creativity.

The inherent non-deterministic nature of LLMs can foster a degree of creativity and diversity in the outputs that deterministic systems may not achieve.
This aspect is particularly valuable in applications requiring innovative solutions or creative content generation, where the variety and uniqueness of the output are more critical than in strictly rule-based or deterministic scenarios.
In other words, the non-determinism implies that making multiple requests to LLMs may increase the chance for developers to receive high quality code and therefore enhance the code generation performance.
For instance, through making five requests in RQ1 with temperature of 1, the candidate that achieves the highest pass rate for a given code problem shows an improvement on average of around 16.13 times (CodeContests), 3.12 times (APPS), and 1.98 times (HumanEval) over the candidate with the lowest pass rate;
it exhibits an overall improvement of 5.21 times (CodeContests), 1.40 times (APPS), and 0.59 times (HumanEval) against its mean performance among five candidates.
Looking deeper into the consistency of the error, we can find that generated code candidates are more likely 
(at least 65.85\%, 73.83\%, and 90.00\% in CodeContests, APPS, and HumanEval)
to share the same error type if all of them fail to pass the test cases.
The most common error types they share are IndexError (46.03\% in CodeContests), IndexError (34.78\% in APPS), and NameError (33.33\% in HumanEval) respectively, under temperature=0.

\subsection{Future work}

 Achieving an optimal balance between determinism and creativity is crucial for enhancing LLMs' effectiveness across a broad spectrum of applications.
Too much determinism could stifle creativity, leading to predictable and monotonous outputs, while excessive non-determinism might compromise the reliability and consistency necessary for applications requiring precise and accurate results.
To address these challenges and strike a balance between determinism and creativity, future research could explore several promising directions:

\textbf{Voting Mechanism}: Implementing a voting mechanism wherein multiple candidates of the model generate outputs, and a consensus approach should be used to select the most appropriate output.
This method can help mitigate the effects of non-determinism by leveraging the collective decision-making process to choose outputs that are both creative and relevant to the task.

\textbf{Repair Loop Driven by LLMs}: Developing techniques for loop repair driven by LLMs can offer a novel approach to addressing non-determinism.
By automatically identifying and correcting inconsistencies or errors in the generated code, such a system could enhance the reliability of outputs without significantly compromising creativity.
This approach would rely on the model's ability to learn from feedback loops, improving its performance over time.

\textbf{Hybrid Models}: Investigating hybrid models that combine deterministic and non-deterministic components might offer a pathway to achieving the desired balance.
Such models could leverage the strengths of both approaches, using deterministic methods to ensure reliability and consistency where needed, while allowing for creative freedom through non-deterministic processes in aspects where innovation is prized.

\textbf{Customizable Levels of Determinism}: Developing LLMs that allow users to specify their preferred level of determinism versus creativity could cater to a wide range of applications.
This customization could enable users to tune the model's outputs according to the specific requirements (e.g. domain-specific) of their task, whether that be generating highly creative content or producing consistent and reliable code.

\section{Conclusion}

This work studies the non-determinism threat of code generation with ChatGPT. 
We perform experiments on three widely studied code generation benchmarks and find that
the correctness, test outputs, as well as syntax and structure of code candidates generated from the same instruction, vary significantly in different requests.
We hope that this paper could raise awareness of the threat of non-determinism in future code generation tasks when using large language models.
%

\section{Acknowledgement}
This work was supported by the UKRI Centre for Doctoral Training in Safe and Trusted Artificial Intelligence (EP/S023356/1).

\bibliographystyle{ACM-Reference-Format}
\bibliography{sample-manuscript}


\begin{thebibliography}{71}


\ifx \showCODEN    \undefined \def \showCODEN     #1{\unskip}     \fi
\ifx \showDOI      \undefined \def \showDOI       #1{#1}\fi
\ifx \showISBNx    \undefined \def \showISBNx     #1{\unskip}     \fi
\ifx \showISBNxiii \undefined \def \showISBNxiii  #1{\unskip}     \fi
\ifx \showISSN     \undefined \def \showISSN      #1{\unskip}     \fi
\ifx \showLCCN     \undefined \def \showLCCN      #1{\unskip}     \fi
\ifx \shownote     \undefined \def \shownote      #1{#1}          \fi
\ifx \showarticletitle \undefined \def \showarticletitle #1{#1}   \fi
\ifx \showURL      \undefined \def \showURL       {\relax}        \fi
\providecommand\bibfield[2]{#2}
\providecommand\bibinfo[2]{#2}
\providecommand\natexlab[1]{#1}
\providecommand\showeprint[2][]{arXiv:#2}

\bibitem[gpt(pt 4)]%
        {gptrandomness}
 \bibinfo{year}{https://152334h.github.io/blog/non-determinism-in-gpt-4/}\natexlab{}.
\newblock
\newblock


\bibitem[ope(chat)]%
        {openaichat}
 \bibinfo{year}{https://chat.openai.com/chat}\natexlab{}.
\newblock
\newblock


\bibitem[hom(late)]%
        {homepage}
 \bibinfo{year}{https://github.com/ShuyinOuyang/LLM-is-a-box-of-chocolate}\natexlab{}.
\newblock
\newblock


\bibitem[Austin et~al\mbox{.}(2021)]%
        {austin2021program}
\bibfield{author}{\bibinfo{person}{Jacob Austin}, \bibinfo{person}{Augustus Odena}, \bibinfo{person}{Maxwell Nye}, \bibinfo{person}{Maarten Bosma}, \bibinfo{person}{Henryk Michalewski}, \bibinfo{person}{David Dohan}, \bibinfo{person}{Ellen Jiang}, \bibinfo{person}{Carrie Cai}, \bibinfo{person}{Michael Terry}, \bibinfo{person}{Quoc Le}, {and} \bibinfo{person}{Charles Sutton}.} \bibinfo{year}{2021}\natexlab{}.
\newblock \bibinfo{title}{Program Synthesis with Large Language Models}.
\newblock
\newblock
\showeprint[arxiv]{2108.07732}~[cs.PL]
\urldef\tempurl%
\url{https://arxiv.org/abs/2108.07732}
\showURL{%
\tempurl}


\bibitem[Baldini et~al\mbox{.}(2022)]%
        {soares2022your}
\bibfield{author}{\bibinfo{person}{Ioana Baldini}, \bibinfo{person}{Dennis Wei}, \bibinfo{person}{Karthikeyan Natesan~Ramamurthy}, \bibinfo{person}{Moninder Singh}, {and} \bibinfo{person}{Mikhail Yurochkin}.} \bibinfo{year}{2022}\natexlab{}.
\newblock \showarticletitle{Your fairness may vary: Pretrained language model fairness in toxic text classification}. In \bibinfo{booktitle}{\emph{Findings of the Association for Computational Linguistics: ACL 2022}}, \bibfield{editor}{\bibinfo{person}{Smaranda Muresan}, \bibinfo{person}{Preslav Nakov}, {and} \bibinfo{person}{Aline Villavicencio}} (Eds.). \bibinfo{publisher}{Association for Computational Linguistics}, \bibinfo{address}{Dublin, Ireland}, \bibinfo{pages}{2245--2262}.
\newblock
\urldef\tempurl%
\url{https://doi.org/10.18653/v1/2022.findings-acl.176}
\showDOI{\tempurl}


\bibitem[Bang et~al\mbox{.}(2023)]%
        {bang2023multitask}
\bibfield{author}{\bibinfo{person}{Y Bang}, \bibinfo{person}{S Cahyawijaya}, \bibinfo{person}{N Lee}, \bibinfo{person}{W Dai}, \bibinfo{person}{D Su}, \bibinfo{person}{B Wilie}, \bibinfo{person}{H Lovenia}, \bibinfo{person}{Z Ji}, \bibinfo{person}{T Yu}, \bibinfo{person}{W Chung}, {et~al\mbox{.}}} \bibinfo{year}{2023}\natexlab{}.
\newblock \bibinfo{title}{A multitask, multilingual, multimodal evaluation of ChatGPT on reasoning, hallucination, and interactivity. arXiv}.
\newblock
\newblock


\bibitem[Bhavya et~al\mbox{.}(2022)]%
        {bhavya2022analogy}
\bibfield{author}{\bibinfo{person}{Bhavya Bhavya}, \bibinfo{person}{Jinjun Xiong}, {and} \bibinfo{person}{Chengxiang Zhai}.} \bibinfo{year}{2022}\natexlab{}.
\newblock \bibinfo{title}{Analogy Generation by Prompting Large Language Models: A Case Study of InstructGPT}.
\newblock
\newblock
\showeprint[arxiv]{2210.04186}~[cs.CL]
\urldef\tempurl%
\url{https://arxiv.org/abs/2210.04186}
\showURL{%
\tempurl}


\bibitem[Biermann(1978)]%
        {biermann1978inference}
\bibfield{author}{\bibinfo{person}{Alan~W Biermann}.} \bibinfo{year}{1978}\natexlab{}.
\newblock \showarticletitle{The inference of regular LISP programs from examples}.
\newblock \bibinfo{journal}{\emph{IEEE transactions on Systems, Man, and Cybernetics}} \bibinfo{volume}{8}, \bibinfo{number}{8} (\bibinfo{year}{1978}), \bibinfo{pages}{585--600}.
\newblock


\bibitem[Brown et~al\mbox{.}(2020)]%
        {brown2020language}
\bibfield{author}{\bibinfo{person}{Tom Brown}, \bibinfo{person}{Benjamin Mann}, \bibinfo{person}{Nick Ryder}, \bibinfo{person}{Melanie Subbiah}, \bibinfo{person}{Jared~D Kaplan}, \bibinfo{person}{Prafulla Dhariwal}, \bibinfo{person}{Arvind Neelakantan}, \bibinfo{person}{Pranav Shyam}, \bibinfo{person}{Girish Sastry}, \bibinfo{person}{Amanda Askell}, {et~al\mbox{.}}} \bibinfo{year}{2020}\natexlab{}.
\newblock \showarticletitle{Language models are few-shot learners}.
\newblock \bibinfo{journal}{\emph{Advances in neural information processing systems}}  \bibinfo{volume}{33} (\bibinfo{year}{2020}), \bibinfo{pages}{1877--1901}.
\newblock


\bibitem[Bubeck et~al\mbox{.}(2023)]%
        {bubeck2023sparks}
\bibfield{author}{\bibinfo{person}{Sébastien Bubeck}, \bibinfo{person}{Varun Chandrasekaran}, \bibinfo{person}{Ronen Eldan}, \bibinfo{person}{Johannes Gehrke}, \bibinfo{person}{Eric Horvitz}, \bibinfo{person}{Ece Kamar}, \bibinfo{person}{Peter Lee}, \bibinfo{person}{Yin~Tat Lee}, \bibinfo{person}{Yuanzhi Li}, \bibinfo{person}{Scott Lundberg}, \bibinfo{person}{Harsha Nori}, \bibinfo{person}{Hamid Palangi}, \bibinfo{person}{Marco~Tulio Ribeiro}, {and} \bibinfo{person}{Yi Zhang}.} \bibinfo{year}{2023}\natexlab{}.
\newblock \bibinfo{title}{Sparks of Artificial General Intelligence: Early experiments with GPT-4}.
\newblock
\newblock
\showeprint[arxiv]{2303.12712}~[cs.CL]
\urldef\tempurl%
\url{https://arxiv.org/abs/2303.12712}
\showURL{%
\tempurl}


\bibitem[Chatterjee et~al\mbox{.}(2022)]%
        {chatterjee2022reliability}
\bibfield{author}{\bibinfo{person}{Subhashis Chatterjee}, \bibinfo{person}{Deepjyoti Saha}, \bibinfo{person}{Akhilesh Sharma}, {and} \bibinfo{person}{Yogesh Verma}.} \bibinfo{year}{2022}\natexlab{}.
\newblock \bibinfo{title}{Reliability and optimal release time analysis for multi up-gradation software with imperfect debugging and varied testing coverage under the effect of random field environments}.
\newblock , \bibinfo{numpages}{21}~pages.
\newblock


\bibitem[Chen et~al\mbox{.}(2021)]%
        {chen2021evaluating}
\bibfield{author}{\bibinfo{person}{Mark Chen}, \bibinfo{person}{Jerry Tworek}, \bibinfo{person}{Heewoo Jun}, \bibinfo{person}{Qiming Yuan}, \bibinfo{person}{Henrique~Ponde de Oliveira~Pinto}, \bibinfo{person}{Jared Kaplan}, \bibinfo{person}{Harri Edwards}, \bibinfo{person}{Yuri Burda}, \bibinfo{person}{Nicholas Joseph}, \bibinfo{person}{Greg Brockman}, \bibinfo{person}{Alex Ray}, \bibinfo{person}{Raul Puri}, \bibinfo{person}{Gretchen Krueger}, \bibinfo{person}{Michael Petrov}, \bibinfo{person}{Heidy Khlaaf}, \bibinfo{person}{Girish Sastry}, \bibinfo{person}{Pamela Mishkin}, \bibinfo{person}{Brooke Chan}, \bibinfo{person}{Scott Gray}, \bibinfo{person}{Nick Ryder}, \bibinfo{person}{Mikhail Pavlov}, \bibinfo{person}{Alethea Power}, \bibinfo{person}{Lukasz Kaiser}, \bibinfo{person}{Mohammad Bavarian}, \bibinfo{person}{Clemens Winter}, \bibinfo{person}{Philippe Tillet}, \bibinfo{person}{Felipe~Petroski Such}, \bibinfo{person}{Dave Cummings}, \bibinfo{person}{Matthias Plappert}, \bibinfo{person}{Fotios Chantzis},
  \bibinfo{person}{Elizabeth Barnes}, \bibinfo{person}{Ariel Herbert-Voss}, \bibinfo{person}{William~Hebgen Guss}, \bibinfo{person}{Alex Nichol}, \bibinfo{person}{Alex Paino}, \bibinfo{person}{Nikolas Tezak}, \bibinfo{person}{Jie Tang}, \bibinfo{person}{Igor Babuschkin}, \bibinfo{person}{Suchir Balaji}, \bibinfo{person}{Shantanu Jain}, \bibinfo{person}{William Saunders}, \bibinfo{person}{Christopher Hesse}, \bibinfo{person}{Andrew~N. Carr}, \bibinfo{person}{Jan Leike}, \bibinfo{person}{Josh Achiam}, \bibinfo{person}{Vedant Misra}, \bibinfo{person}{Evan Morikawa}, \bibinfo{person}{Alec Radford}, \bibinfo{person}{Matthew Knight}, \bibinfo{person}{Miles Brundage}, \bibinfo{person}{Mira Murati}, \bibinfo{person}{Katie Mayer}, \bibinfo{person}{Peter Welinder}, \bibinfo{person}{Bob McGrew}, \bibinfo{person}{Dario Amodei}, \bibinfo{person}{Sam McCandlish}, \bibinfo{person}{Ilya Sutskever}, {and} \bibinfo{person}{Wojciech Zaremba}.} \bibinfo{year}{2021}\natexlab{}.
\newblock \bibinfo{title}{Evaluating Large Language Models Trained on Code}.
\newblock
\newblock
\showeprint[arxiv]{2107.03374}~[cs.LG]
\urldef\tempurl%
\url{https://arxiv.org/abs/2107.03374}
\showURL{%
\tempurl}


\bibitem[Deng et~al\mbox{.}(2024)]%
        {deng2023large}
\bibfield{author}{\bibinfo{person}{Yinlin Deng}, \bibinfo{person}{Chunqiu~Steven Xia}, \bibinfo{person}{Chenyuan Yang}, \bibinfo{person}{Shizhuo~Dylan Zhang}, \bibinfo{person}{Shujing Yang}, {and} \bibinfo{person}{Lingming Zhang}.} \bibinfo{year}{2024}\natexlab{}.
\newblock \showarticletitle{Large Language Models are Edge-Case Generators: Crafting Unusual Programs for Fuzzing Deep Learning Libraries}. In \bibinfo{booktitle}{\emph{Proceedings of the IEEE/ACM 46th International Conference on Software Engineering}} (Lisbon, Portugal) \emph{(\bibinfo{series}{ICSE '24})}. \bibinfo{publisher}{Association for Computing Machinery}, \bibinfo{address}{New York, NY, USA}, Article \bibinfo{articleno}{70}, \bibinfo{numpages}{13}~pages.
\newblock
\showISBNx{9798400702174}
\urldef\tempurl%
\url{https://doi.org/10.1145/3597503.3623343}
\showDOI{\tempurl}


\bibitem[Dong and Lapata(2016)]%
        {dong2016language}
\bibfield{author}{\bibinfo{person}{Li Dong} {and} \bibinfo{person}{Mirella Lapata}.} \bibinfo{year}{2016}\natexlab{}.
\newblock \showarticletitle{Language to Logical Form with Neural Attention}. In \bibinfo{booktitle}{\emph{Proceedings of the 54th Annual Meeting of the Association for Computational Linguistics (Volume 1: Long Papers)}}, \bibfield{editor}{\bibinfo{person}{Katrin Erk} {and} \bibinfo{person}{Noah~A. Smith}} (Eds.). \bibinfo{publisher}{Association for Computational Linguistics}, \bibinfo{address}{Berlin, Germany}, \bibinfo{pages}{33--43}.
\newblock
\urldef\tempurl%
\url{https://doi.org/10.18653/v1/P16-1004}
\showDOI{\tempurl}


\bibitem[Fan et~al\mbox{.}(2023)]%
        {fan2023large}
\bibfield{author}{\bibinfo{person}{Angela Fan}, \bibinfo{person}{Beliz Gokkaya}, \bibinfo{person}{Mark Harman}, \bibinfo{person}{Mitya Lyubarskiy}, \bibinfo{person}{Shubho Sengupta}, \bibinfo{person}{Shin Yoo}, {and} \bibinfo{person}{Jie~M. Zhang}.} \bibinfo{year}{2023}\natexlab{}.
\newblock \bibinfo{title}{Large Language Models for Software Engineering: Survey and Open Problems}.
\newblock
\newblock
\showeprint[arxiv]{2310.03533}~[cs.SE]
\urldef\tempurl%
\url{https://arxiv.org/abs/2310.03533}
\showURL{%
\tempurl}


\bibitem[Feng et~al\mbox{.}(2023)]%
        {feng2023investigating}
\bibfield{author}{\bibinfo{person}{Yunhe Feng}, \bibinfo{person}{Sreecharan Vanam}, \bibinfo{person}{Manasa Cherukupally}, \bibinfo{person}{Weijian Zheng}, \bibinfo{person}{Meikang Qiu}, {and} \bibinfo{person}{Haihua Chen}.} \bibinfo{year}{2023}\natexlab{}.
\newblock \showarticletitle{Investigating Code Generation Performance of ChatGPT with Crowdsourcing Social Data}. In \bibinfo{booktitle}{\emph{2023 IEEE 47th Annual Computers, Software, and Applications Conference (COMPSAC)}}. \bibinfo{publisher}{IEEE Computer Society}, \bibinfo{address}{Los Alamitos, CA, USA}, \bibinfo{pages}{876--885}.
\newblock
\showISSN{0730-3157}
\urldef\tempurl%
\url{https://doi.org/10.1109/COMPSAC57700.2023.00117}
\showDOI{\tempurl}


\bibitem[Feng et~al\mbox{.}(2020)]%
        {feng2020codebert}
\bibfield{author}{\bibinfo{person}{Zhangyin Feng}, \bibinfo{person}{Daya Guo}, \bibinfo{person}{Duyu Tang}, \bibinfo{person}{Nan Duan}, \bibinfo{person}{Xiaocheng Feng}, \bibinfo{person}{Ming Gong}, \bibinfo{person}{Linjun Shou}, \bibinfo{person}{Bing Qin}, \bibinfo{person}{Ting Liu}, \bibinfo{person}{Daxin Jiang}, {and} \bibinfo{person}{Ming Zhou}.} \bibinfo{year}{2020}\natexlab{}.
\newblock \bibinfo{title}{CodeBERT: A Pre-Trained Model for Programming and Natural Languages}.
\newblock
\newblock
\showeprint[arxiv]{2002.08155}~[cs.CL]
\urldef\tempurl%
\url{https://arxiv.org/abs/2002.08155}
\showURL{%
\tempurl}


\bibitem[Frieder et~al\mbox{.}(2024)]%
        {frieder2024mathematical}
\bibfield{author}{\bibinfo{person}{Simon Frieder}, \bibinfo{person}{Luca Pinchetti}, \bibinfo{person}{Alexis Chevalier}, \bibinfo{person}{Ryan-Rhys Griffiths}, \bibinfo{person}{Tommaso Salvatori}, \bibinfo{person}{Thomas Lukasiewicz}, \bibinfo{person}{Philipp Petersen}, {and} \bibinfo{person}{Julius Berner}.} \bibinfo{year}{2024}\natexlab{}.
\newblock \showarticletitle{Mathematical capabilities of ChatGPT}. In \bibinfo{booktitle}{\emph{Proceedings of the 37th International Conference on Neural Information Processing Systems}} (New Orleans, LA, USA) \emph{(\bibinfo{series}{NIPS '23})}. \bibinfo{publisher}{Curran Associates Inc.}, \bibinfo{address}{Red Hook, NY, USA}, Article \bibinfo{articleno}{1205}, \bibinfo{numpages}{46}~pages.
\newblock


\bibitem[Green(1969)]%
        {green1981application}
\bibfield{author}{\bibinfo{person}{Cordell Green}.} \bibinfo{year}{1969}\natexlab{}.
\newblock \showarticletitle{Application of theorem proving to problem solving}. In \bibinfo{booktitle}{\emph{Proceedings of the 1st International Joint Conference on Artificial Intelligence}} (Washington, DC) \emph{(\bibinfo{series}{IJCAI'69})}. \bibinfo{publisher}{Morgan Kaufmann Publishers Inc.}, \bibinfo{address}{San Francisco, CA, USA}, \bibinfo{pages}{219–239}.
\newblock


\bibitem[Gulwani(2011)]%
        {gulwani2011automating}
\bibfield{author}{\bibinfo{person}{Sumit Gulwani}.} \bibinfo{year}{2011}\natexlab{}.
\newblock \showarticletitle{Automating string processing in spreadsheets using input-output examples}.
\newblock \bibinfo{journal}{\emph{ACM Sigplan Notices}} \bibinfo{volume}{46}, \bibinfo{number}{1} (\bibinfo{year}{2011}), \bibinfo{pages}{317--330}.
\newblock


\bibitem[Gulwani et~al\mbox{.}(2017)]%
        {gulwani2017program}
\bibfield{author}{\bibinfo{person}{Sumit Gulwani}, \bibinfo{person}{Oleksandr Polozov}, \bibinfo{person}{Rishabh Singh}, {et~al\mbox{.}}} \bibinfo{year}{2017}\natexlab{}.
\newblock \showarticletitle{Program synthesis}.
\newblock \bibinfo{journal}{\emph{Foundations and Trends{\textregistered} in Programming Languages}} \bibinfo{volume}{4}, \bibinfo{number}{1-2} (\bibinfo{year}{2017}), \bibinfo{pages}{1--119}.
\newblock


\bibitem[Guo et~al\mbox{.}(2021)]%
        {guo2020graphcodebert}
\bibfield{author}{\bibinfo{person}{Daya Guo}, \bibinfo{person}{Shuo Ren}, \bibinfo{person}{Shuai Lu}, \bibinfo{person}{Zhangyin Feng}, \bibinfo{person}{Duyu Tang}, \bibinfo{person}{Shujie Liu}, \bibinfo{person}{Long Zhou}, \bibinfo{person}{Nan Duan}, \bibinfo{person}{Alexey Svyatkovskiy}, \bibinfo{person}{Shengyu Fu}, \bibinfo{person}{Michele Tufano}, \bibinfo{person}{Shao~Kun Deng}, \bibinfo{person}{Colin Clement}, \bibinfo{person}{Dawn Drain}, \bibinfo{person}{Neel Sundaresan}, \bibinfo{person}{Jian Yin}, \bibinfo{person}{Daxin Jiang}, {and} \bibinfo{person}{Ming Zhou}.} \bibinfo{year}{2021}\natexlab{}.
\newblock \bibinfo{title}{GraphCodeBERT: Pre-training Code Representations with Data Flow}.
\newblock
\newblock
\showeprint[arxiv]{2009.08366}~[cs.SE]
\urldef\tempurl%
\url{https://arxiv.org/abs/2009.08366}
\showURL{%
\tempurl}


\bibitem[Guo et~al\mbox{.}(2023)]%
        {guo2024exploring}
\bibfield{author}{\bibinfo{person}{Qi Guo}, \bibinfo{person}{Junming Cao}, \bibinfo{person}{Xiaofei Xie}, \bibinfo{person}{Shangqing Liu}, \bibinfo{person}{Xiaohong Li}, \bibinfo{person}{Bihuan Chen}, {and} \bibinfo{person}{Xin Peng}.} \bibinfo{year}{2023}\natexlab{}.
\newblock \bibinfo{title}{Exploring the Potential of ChatGPT in Automated Code Refinement: An Empirical Study}.
\newblock
\newblock
\showeprint[arxiv]{2309.08221}~[cs.SE]
\urldef\tempurl%
\url{https://arxiv.org/abs/2309.08221}
\showURL{%
\tempurl}


\bibitem[Hashimoto et~al\mbox{.}(2018)]%
        {hashimoto2018retrieve}
\bibfield{author}{\bibinfo{person}{Tatsunori~B. Hashimoto}, \bibinfo{person}{Kelvin Guu}, \bibinfo{person}{Yonatan Oren}, {and} \bibinfo{person}{Percy Liang}.} \bibinfo{year}{2018}\natexlab{}.
\newblock \showarticletitle{A retrieve-and-edit framework for predicting structured outputs}. In \bibinfo{booktitle}{\emph{Proceedings of the 32nd International Conference on Neural Information Processing Systems}} (Montr\'{e}al, Canada) \emph{(\bibinfo{series}{NIPS'18})}. \bibinfo{publisher}{Curran Associates Inc.}, \bibinfo{address}{Red Hook, NY, USA}, \bibinfo{pages}{10073–10083}.
\newblock


\bibitem[Hassani and Silva(2023)]%
        {hassani2023role}
\bibfield{author}{\bibinfo{person}{Hossein Hassani} {and} \bibinfo{person}{Emmanuel~Sirmal Silva}.} \bibinfo{year}{2023}\natexlab{}.
\newblock \showarticletitle{The role of ChatGPT in data science: how ai-assisted conversational interfaces are revolutionizing the field}.
\newblock \bibinfo{journal}{\emph{Big data and cognitive computing}} \bibinfo{volume}{7}, \bibinfo{number}{2} (\bibinfo{year}{2023}), \bibinfo{pages}{62}.
\newblock


\bibitem[Hendrycks et~al\mbox{.}(2021)]%
        {hendrycks2021measuring}
\bibfield{author}{\bibinfo{person}{Dan Hendrycks}, \bibinfo{person}{Steven Basart}, \bibinfo{person}{Saurav Kadavath}, \bibinfo{person}{Mantas Mazeika}, \bibinfo{person}{Akul Arora}, \bibinfo{person}{Ethan Guo}, \bibinfo{person}{Collin Burns}, \bibinfo{person}{Samir Puranik}, \bibinfo{person}{Horace He}, \bibinfo{person}{Dawn Song}, {and} \bibinfo{person}{Jacob Steinhardt}.} \bibinfo{year}{2021}\natexlab{}.
\newblock \bibinfo{title}{Measuring Coding Challenge Competence With APPS}.
\newblock
\newblock
\showeprint[arxiv]{2105.09938}~[cs.SE]
\urldef\tempurl%
\url{https://arxiv.org/abs/2105.09938}
\showURL{%
\tempurl}


\bibitem[Hendrycks et~al\mbox{.}(2023)]%
        {hendrycks2020aligning}
\bibfield{author}{\bibinfo{person}{Dan Hendrycks}, \bibinfo{person}{Collin Burns}, \bibinfo{person}{Steven Basart}, \bibinfo{person}{Andrew Critch}, \bibinfo{person}{Jerry Li}, \bibinfo{person}{Dawn Song}, {and} \bibinfo{person}{Jacob Steinhardt}.} \bibinfo{year}{2023}\natexlab{}.
\newblock \bibinfo{title}{Aligning AI With Shared Human Values}.
\newblock
\newblock
\showeprint[arxiv]{2008.02275}~[cs.CY]
\urldef\tempurl%
\url{https://arxiv.org/abs/2008.02275}
\showURL{%
\tempurl}


\bibitem[Inala et~al\mbox{.}(2022)]%
        {inala2022fault}
\bibfield{author}{\bibinfo{person}{Jeevana~Priya Inala}, \bibinfo{person}{Chenglong Wang}, \bibinfo{person}{Mei Yang}, \bibinfo{person}{Andres Codas}, \bibinfo{person}{Mark Encarnaci{\'o}n}, \bibinfo{person}{Shuvendu Lahiri}, \bibinfo{person}{Madanlal Musuvathi}, {and} \bibinfo{person}{Jianfeng Gao}.} \bibinfo{year}{2022}\natexlab{}.
\newblock \showarticletitle{Fault-aware neural code rankers}.
\newblock \bibinfo{journal}{\emph{Advances in Neural Information Processing Systems}}  \bibinfo{volume}{35} (\bibinfo{year}{2022}), \bibinfo{pages}{13419--13432}.
\newblock


\bibitem[Jalil et~al\mbox{.}(2023)]%
        {jalil2023chatgpt}
\bibfield{author}{\bibinfo{person}{Sajed Jalil}, \bibinfo{person}{Suzzana Rafi}, \bibinfo{person}{Thomas~D. LaToza}, \bibinfo{person}{Kevin Moran}, {and} \bibinfo{person}{Wing Lam}.} \bibinfo{year}{2023}\natexlab{}.
\newblock \bibinfo{title}{ChatGPT and Software Testing Education: Promises \& Perils}.
\newblock
\newblock
\urldef\tempurl%
\url{https://doi.org/10.1109/icstw58534.2023.00078}
\showDOI{\tempurl}


\bibitem[Kiviriga(2023)]%
        {kiviriga2023efficient}
\bibfield{author}{\bibinfo{person}{Andrej Kiviriga}.} \bibinfo{year}{2023}\natexlab{}.
\newblock \bibinfo{title}{Efficient Model Checking: The Power of Randomness}.
\newblock
\newblock


\bibitem[Krishna et~al\mbox{.}(2022)]%
        {krishna2022rankgen}
\bibfield{author}{\bibinfo{person}{Kalpesh Krishna}, \bibinfo{person}{Yapei Chang}, \bibinfo{person}{John Wieting}, {and} \bibinfo{person}{Mohit Iyyer}.} \bibinfo{year}{2022}\natexlab{}.
\newblock \bibinfo{title}{RankGen: Improving Text Generation with Large Ranking Models}.
\newblock
\newblock
\showeprint[arxiv]{2205.09726}~[cs.CL]
\urldef\tempurl%
\url{https://arxiv.org/abs/2205.09726}
\showURL{%
\tempurl}


\bibitem[Kulal et~al\mbox{.}(2019)]%
        {kulal2019spoc}
\bibfield{author}{\bibinfo{person}{Sumith Kulal}, \bibinfo{person}{Panupong Pasupat}, \bibinfo{person}{Kartik Chandra}, \bibinfo{person}{Mina Lee}, \bibinfo{person}{Oded Padon}, \bibinfo{person}{Alex Aiken}, {and} \bibinfo{person}{Percy Liang}.} \bibinfo{year}{2019}\natexlab{}.
\newblock \bibinfo{title}{SPoC: Search-based Pseudocode to Code}.
\newblock
\newblock
\showeprint[arxiv]{1906.04908}~[cs.LG]
\urldef\tempurl%
\url{https://arxiv.org/abs/1906.04908}
\showURL{%
\tempurl}


\bibitem[Lee et~al\mbox{.}(2022)]%
        {lee2022coauthor}
\bibfield{author}{\bibinfo{person}{Mina Lee}, \bibinfo{person}{Percy Liang}, {and} \bibinfo{person}{Qian Yang}.} \bibinfo{year}{2022}\natexlab{}.
\newblock \showarticletitle{CoAuthor: Designing a Human-AI Collaborative Writing Dataset for Exploring Language Model Capabilities}. In \bibinfo{booktitle}{\emph{Proceedings of the 2022 CHI Conference on Human Factors in Computing Systems}} (New Orleans, LA, USA) \emph{(\bibinfo{series}{CHI '22})}. \bibinfo{publisher}{Association for Computing Machinery}, \bibinfo{address}{New York, NY, USA}, Article \bibinfo{articleno}{388}, \bibinfo{numpages}{19}~pages.
\newblock
\showISBNx{9781450391573}
\urldef\tempurl%
\url{https://doi.org/10.1145/3491102.3502030}
\showDOI{\tempurl}


\bibitem[Li et~al\mbox{.}(2023b)]%
        {li2023codeeditor}
\bibfield{author}{\bibinfo{person}{Jia Li}, \bibinfo{person}{Ge Li}, \bibinfo{person}{Zhuo Li}, \bibinfo{person}{Zhi Jin}, \bibinfo{person}{Xing Hu}, \bibinfo{person}{Kechi Zhang}, {and} \bibinfo{person}{Zhiyi Fu}.} \bibinfo{year}{2023}\natexlab{b}.
\newblock \showarticletitle{Codeeditor: Learning to edit source code with pre-trained models}.
\newblock \bibinfo{journal}{\emph{ACM Transactions on Software Engineering and Methodology}} \bibinfo{volume}{32}, \bibinfo{number}{6} (\bibinfo{year}{2023}), \bibinfo{pages}{1--22}.
\newblock


\bibitem[Li et~al\mbox{.}(2023a)]%
        {li2023skcoder}
\bibfield{author}{\bibinfo{person}{Jia Li}, \bibinfo{person}{Yongmin Li}, \bibinfo{person}{Ge Li}, \bibinfo{person}{Zhi Jin}, \bibinfo{person}{Yiyang Hao}, {and} \bibinfo{person}{Xing Hu}.} \bibinfo{year}{2023}\natexlab{a}.
\newblock \bibinfo{title}{SkCoder: A Sketch-based Approach for Automatic Code Generation}.
\newblock
\newblock
\showeprint[arxiv]{2302.06144}~[cs.SE]
\urldef\tempurl%
\url{https://arxiv.org/abs/2302.06144}
\showURL{%
\tempurl}


\bibitem[Li et~al\mbox{.}(2022)]%
        {li2022competition}
\bibfield{author}{\bibinfo{person}{Yujia Li}, \bibinfo{person}{David Choi}, \bibinfo{person}{Junyoung Chung}, \bibinfo{person}{Nate Kushman}, \bibinfo{person}{Julian Schrittwieser}, \bibinfo{person}{R{\'e}mi Leblond}, \bibinfo{person}{Tom Eccles}, \bibinfo{person}{James Keeling}, \bibinfo{person}{Felix Gimeno}, \bibinfo{person}{Agustin Dal~Lago}, {et~al\mbox{.}}} \bibinfo{year}{2022}\natexlab{}.
\newblock \showarticletitle{Competition-level code generation with alphacode}.
\newblock \bibinfo{journal}{\emph{Science}} \bibinfo{volume}{378}, \bibinfo{number}{6624} (\bibinfo{year}{2022}), \bibinfo{pages}{1092--1097}.
\newblock


\bibitem[Li et~al\mbox{.}(2023c)]%
        {li2023cctest}
\bibfield{author}{\bibinfo{person}{Zongjie Li}, \bibinfo{person}{Chaozheng Wang}, \bibinfo{person}{Zhibo Liu}, \bibinfo{person}{Haoxuan Wang}, \bibinfo{person}{Dong Chen}, \bibinfo{person}{Shuai Wang}, {and} \bibinfo{person}{Cuiyun Gao}.} \bibinfo{year}{2023}\natexlab{c}.
\newblock \bibinfo{title}{CCTEST: Testing and Repairing Code Completion Systems}.
\newblock
\newblock
\showeprint[arxiv]{2208.08289}~[cs.SE]
\urldef\tempurl%
\url{https://arxiv.org/abs/2208.08289}
\showURL{%
\tempurl}


\bibitem[Liang et~al\mbox{.}(2023)]%
        {liang2023code}
\bibfield{author}{\bibinfo{person}{Jacky Liang}, \bibinfo{person}{Wenlong Huang}, \bibinfo{person}{Fei Xia}, \bibinfo{person}{Peng Xu}, \bibinfo{person}{Karol Hausman}, \bibinfo{person}{Brian Ichter}, \bibinfo{person}{Pete Florence}, {and} \bibinfo{person}{Andy Zeng}.} \bibinfo{year}{2023}\natexlab{}.
\newblock \bibinfo{title}{Code as Policies: Language Model Programs for Embodied Control}.
\newblock
\newblock
\showeprint[arxiv]{2209.07753}~[cs.RO]
\urldef\tempurl%
\url{https://arxiv.org/abs/2209.07753}
\showURL{%
\tempurl}


\bibitem[Ling et~al\mbox{.}(2016)]%
        {ling2016latent}
\bibfield{author}{\bibinfo{person}{Wang Ling}, \bibinfo{person}{Edward Grefenstette}, \bibinfo{person}{Karl~Moritz Hermann}, \bibinfo{person}{Tomáš Kočiský}, \bibinfo{person}{Andrew Senior}, \bibinfo{person}{Fumin Wang}, {and} \bibinfo{person}{Phil Blunsom}.} \bibinfo{year}{2016}\natexlab{}.
\newblock \bibinfo{title}{Latent Predictor Networks for Code Generation}.
\newblock
\newblock
\showeprint[arxiv]{1603.06744}~[cs.CL]
\urldef\tempurl%
\url{https://arxiv.org/abs/1603.06744}
\showURL{%
\tempurl}


\bibitem[Liu et~al\mbox{.}(2023b)]%
        {liu2023your}
\bibfield{author}{\bibinfo{person}{Jiawei Liu}, \bibinfo{person}{Chunqiu~Steven Xia}, \bibinfo{person}{Yuyao Wang}, {and} \bibinfo{person}{Lingming Zhang}.} \bibinfo{year}{2023}\natexlab{b}.
\newblock \bibinfo{title}{Is Your Code Generated by ChatGPT Really Correct? Rigorous Evaluation of Large Language Models for Code Generation}.
\newblock
\newblock
\showeprint[arxiv]{2305.01210}~[cs.SE]
\urldef\tempurl%
\url{https://arxiv.org/abs/2305.01210}
\showURL{%
\tempurl}


\bibitem[Liu et~al\mbox{.}(2023a)]%
        {liu2023summary}
\bibfield{author}{\bibinfo{person}{Yiheng Liu}, \bibinfo{person}{Tianle Han}, \bibinfo{person}{Siyuan Ma}, \bibinfo{person}{Jiayue Zhang}, \bibinfo{person}{Yuanyuan Yang}, \bibinfo{person}{Jiaming Tian}, \bibinfo{person}{Hao He}, \bibinfo{person}{Antong Li}, \bibinfo{person}{Mengshen He}, \bibinfo{person}{Zhengliang Liu}, \bibinfo{person}{Zihao Wu}, \bibinfo{person}{Lin Zhao}, \bibinfo{person}{Dajiang Zhu}, \bibinfo{person}{Xiang Li}, \bibinfo{person}{Ning Qiang}, \bibinfo{person}{Dingang Shen}, \bibinfo{person}{Tianming Liu}, {and} \bibinfo{person}{Bao Ge}.} \bibinfo{year}{2023}\natexlab{a}.
\newblock \showarticletitle{Summary of ChatGPT-Related research and perspective towards the future of large language models}.
\newblock \bibinfo{journal}{\emph{Meta-Radiology}} \bibinfo{volume}{1}, \bibinfo{number}{2} (\bibinfo{date}{Sept.} \bibinfo{year}{2023}), \bibinfo{pages}{100017}.
\newblock
\showISSN{2950-1628}
\urldef\tempurl%
\url{https://doi.org/10.1016/j.metrad.2023.100017}
\showDOI{\tempurl}


\bibitem[Liu et~al\mbox{.}(2024)]%
        {liu2024refining}
\bibfield{author}{\bibinfo{person}{Yue Liu}, \bibinfo{person}{Thanh Le-Cong}, \bibinfo{person}{Ratnadira Widyasari}, \bibinfo{person}{Chakkrit Tantithamthavorn}, \bibinfo{person}{Li Li}, \bibinfo{person}{Xuan-Bach~D Le}, {and} \bibinfo{person}{David Lo}.} \bibinfo{year}{2024}\natexlab{}.
\newblock \showarticletitle{Refining chatgpt-generated code: Characterizing and mitigating code quality issues}.
\newblock \bibinfo{journal}{\emph{ACM Transactions on Software Engineering and Methodology}} \bibinfo{volume}{33}, \bibinfo{number}{5} (\bibinfo{year}{2024}), \bibinfo{pages}{1--26}.
\newblock


\bibitem[Lu et~al\mbox{.}(2021)]%
        {lu2021codexglue}
\bibfield{author}{\bibinfo{person}{Shuai Lu}, \bibinfo{person}{Daya Guo}, \bibinfo{person}{Shuo Ren}, \bibinfo{person}{Junjie Huang}, \bibinfo{person}{Alexey Svyatkovskiy}, \bibinfo{person}{Ambrosio Blanco}, \bibinfo{person}{Colin Clement}, \bibinfo{person}{Dawn Drain}, \bibinfo{person}{Daxin Jiang}, \bibinfo{person}{Duyu Tang}, \bibinfo{person}{Ge Li}, \bibinfo{person}{Lidong Zhou}, \bibinfo{person}{Linjun Shou}, \bibinfo{person}{Long Zhou}, \bibinfo{person}{Michele Tufano}, \bibinfo{person}{Ming Gong}, \bibinfo{person}{Ming Zhou}, \bibinfo{person}{Nan Duan}, \bibinfo{person}{Neel Sundaresan}, \bibinfo{person}{Shao~Kun Deng}, \bibinfo{person}{Shengyu Fu}, {and} \bibinfo{person}{Shujie Liu}.} \bibinfo{year}{2021}\natexlab{}.
\newblock \bibinfo{title}{CodeXGLUE: A Machine Learning Benchmark Dataset for Code Understanding and Generation}.
\newblock
\newblock
\showeprint[arxiv]{2102.04664}~[cs.SE]
\urldef\tempurl%
\url{https://arxiv.org/abs/2102.04664}
\showURL{%
\tempurl}


\bibitem[Malfa et~al\mbox{.}(2023)]%
        {la2023arrt}
\bibfield{author}{\bibinfo{person}{Emanuele~La Malfa}, \bibinfo{person}{Aleksandar Petrov}, \bibinfo{person}{Simon Frieder}, \bibinfo{person}{Christoph Weinhuber}, \bibinfo{person}{Ryan Burnell}, \bibinfo{person}{Raza Nazar}, \bibinfo{person}{Anthony~G. Cohn}, \bibinfo{person}{Nigel Shadbolt}, {and} \bibinfo{person}{Michael Wooldridge}.} \bibinfo{year}{2023}\natexlab{}.
\newblock \bibinfo{title}{Language Models as a Service: Overview of a New Paradigm and its Challenges}.
\newblock
\newblock
\showeprint[arxiv]{2309.16573}~[cs.AI]
\urldef\tempurl%
\url{https://arxiv.org/abs/2309.16573}
\showURL{%
\tempurl}


\bibitem[Manna and Waldinger(1971)]%
        {manna1971toward}
\bibfield{author}{\bibinfo{person}{Zohar Manna} {and} \bibinfo{person}{Richard~J Waldinger}.} \bibinfo{year}{1971}\natexlab{}.
\newblock \showarticletitle{Toward automatic program synthesis}.
\newblock \bibinfo{journal}{\emph{Commun. ACM}} \bibinfo{volume}{14}, \bibinfo{number}{3} (\bibinfo{year}{1971}), \bibinfo{pages}{151--165}.
\newblock


\bibitem[Mastropaolo et~al\mbox{.}(2023)]%
        {mastropaolo2023robustness}
\bibfield{author}{\bibinfo{person}{Antonio Mastropaolo}, \bibinfo{person}{Luca Pascarella}, \bibinfo{person}{Emanuela Guglielmi}, \bibinfo{person}{Matteo Ciniselli}, \bibinfo{person}{Simone Scalabrino}, \bibinfo{person}{Rocco Oliveto}, {and} \bibinfo{person}{Gabriele Bavota}.} \bibinfo{year}{2023}\natexlab{}.
\newblock \bibinfo{title}{On the Robustness of Code Generation Techniques: An Empirical Study on GitHub Copilot}.
\newblock
\newblock
\showeprint[arxiv]{2302.00438}~[cs.SE]
\urldef\tempurl%
\url{https://arxiv.org/abs/2302.00438}
\showURL{%
\tempurl}


\bibitem[McKight and Najab(2010)]%
        {mckight2010kruskal}
\bibfield{author}{\bibinfo{person}{Patrick~E McKight} {and} \bibinfo{person}{Julius Najab}.} \bibinfo{year}{2010}\natexlab{}.
\newblock \bibinfo{title}{Kruskal-wallis test}.
\newblock , \bibinfo{numpages}{1}~pages.
\newblock


\bibitem[McKnight and Najab(2010)]%
        {mcknight2010mann}
\bibfield{author}{\bibinfo{person}{Patrick~E McKnight} {and} \bibinfo{person}{Julius Najab}.} \bibinfo{year}{2010}\natexlab{}.
\newblock \bibinfo{title}{Mann-Whitney U Test}.
\newblock , \bibinfo{numpages}{1}~pages.
\newblock


\bibitem[Mitchell et~al\mbox{.}(2023)]%
        {mitchell2023detectgpt}
\bibfield{author}{\bibinfo{person}{Eric Mitchell}, \bibinfo{person}{Yoonho Lee}, \bibinfo{person}{Alexander Khazatsky}, \bibinfo{person}{Christopher~D. Manning}, {and} \bibinfo{person}{Chelsea Finn}.} \bibinfo{year}{2023}\natexlab{}.
\newblock \bibinfo{title}{DetectGPT: Zero-Shot Machine-Generated Text Detection using Probability Curvature}.
\newblock
\newblock
\showeprint[arxiv]{2301.11305}~[cs.CL]
\urldef\tempurl%
\url{https://arxiv.org/abs/2301.11305}
\showURL{%
\tempurl}


\bibitem[Nagarajan et~al\mbox{.}(2019)]%
        {nagarajan2018deterministic}
\bibfield{author}{\bibinfo{person}{Prabhat Nagarajan}, \bibinfo{person}{Garrett Warnell}, {and} \bibinfo{person}{Peter Stone}.} \bibinfo{year}{2019}\natexlab{}.
\newblock \bibinfo{title}{Deterministic Implementations for Reproducibility in Deep Reinforcement Learning}.
\newblock
\newblock
\showeprint[arxiv]{1809.05676}~[cs.AI]
\urldef\tempurl%
\url{https://arxiv.org/abs/1809.05676}
\showURL{%
\tempurl}


\bibitem[OpenAI et~al\mbox{.}(2024)]%
        {openai2023gpt4}
\bibfield{author}{\bibinfo{person}{OpenAI}, \bibinfo{person}{Josh Achiam}, \bibinfo{person}{Steven Adler}, \bibinfo{person}{Sandhini Agarwal}, \bibinfo{person}{Lama Ahmad}, \bibinfo{person}{Ilge Akkaya}, \bibinfo{person}{Florencia~Leoni Aleman}, \bibinfo{person}{Diogo Almeida}, \bibinfo{person}{Janko Altenschmidt}, \bibinfo{person}{Sam Altman}, \bibinfo{person}{Shyamal Anadkat}, \bibinfo{person}{Red Avila}, \bibinfo{person}{Igor Babuschkin}, \bibinfo{person}{Suchir Balaji}, \bibinfo{person}{Valerie Balcom}, \bibinfo{person}{Paul Baltescu}, \bibinfo{person}{Haiming Bao}, \bibinfo{person}{Mohammad Bavarian}, \bibinfo{person}{Jeff Belgum}, \bibinfo{person}{Irwan Bello}, \bibinfo{person}{Jake Berdine}, \bibinfo{person}{Gabriel Bernadett-Shapiro}, \bibinfo{person}{Christopher Berner}, \bibinfo{person}{Lenny Bogdonoff}, \bibinfo{person}{Oleg Boiko}, \bibinfo{person}{Madelaine Boyd}, \bibinfo{person}{Anna-Luisa Brakman}, \bibinfo{person}{Greg Brockman}, \bibinfo{person}{Tim Brooks}, \bibinfo{person}{Miles Brundage},
  \bibinfo{person}{Kevin Button}, \bibinfo{person}{Trevor Cai}, \bibinfo{person}{Rosie Campbell}, \bibinfo{person}{Andrew Cann}, \bibinfo{person}{Brittany Carey}, \bibinfo{person}{Chelsea Carlson}, \bibinfo{person}{Rory Carmichael}, \bibinfo{person}{Brooke Chan}, \bibinfo{person}{Che Chang}, \bibinfo{person}{Fotis Chantzis}, \bibinfo{person}{Derek Chen}, \bibinfo{person}{Sully Chen}, \bibinfo{person}{Ruby Chen}, \bibinfo{person}{Jason Chen}, \bibinfo{person}{Mark Chen}, \bibinfo{person}{Ben Chess}, \bibinfo{person}{Chester Cho}, \bibinfo{person}{Casey Chu}, \bibinfo{person}{Hyung~Won Chung}, \bibinfo{person}{Dave Cummings}, \bibinfo{person}{Jeremiah Currier}, \bibinfo{person}{Yunxing Dai}, \bibinfo{person}{Cory Decareaux}, \bibinfo{person}{Thomas Degry}, \bibinfo{person}{Noah Deutsch}, \bibinfo{person}{Damien Deville}, \bibinfo{person}{Arka Dhar}, \bibinfo{person}{David Dohan}, \bibinfo{person}{Steve Dowling}, \bibinfo{person}{Sheila Dunning}, \bibinfo{person}{Adrien Ecoffet}, \bibinfo{person}{Atty Eleti},
  \bibinfo{person}{Tyna Eloundou}, \bibinfo{person}{David Farhi}, \bibinfo{person}{Liam Fedus}, \bibinfo{person}{Niko Felix}, \bibinfo{person}{Simón~Posada Fishman}, \bibinfo{person}{Juston Forte}, \bibinfo{person}{Isabella Fulford}, \bibinfo{person}{Leo Gao}, \bibinfo{person}{Elie Georges}, \bibinfo{person}{Christian Gibson}, \bibinfo{person}{Vik Goel}, \bibinfo{person}{Tarun Gogineni}, \bibinfo{person}{Gabriel Goh}, \bibinfo{person}{Rapha Gontijo-Lopes}, \bibinfo{person}{Jonathan Gordon}, \bibinfo{person}{Morgan Grafstein}, \bibinfo{person}{Scott Gray}, \bibinfo{person}{Ryan Greene}, \bibinfo{person}{Joshua Gross}, \bibinfo{person}{Shixiang~Shane Gu}, \bibinfo{person}{Yufei Guo}, \bibinfo{person}{Chris Hallacy}, \bibinfo{person}{Jesse Han}, \bibinfo{person}{Jeff Harris}, \bibinfo{person}{Yuchen He}, \bibinfo{person}{Mike Heaton}, \bibinfo{person}{Johannes Heidecke}, \bibinfo{person}{Chris Hesse}, \bibinfo{person}{Alan Hickey}, \bibinfo{person}{Wade Hickey}, \bibinfo{person}{Peter Hoeschele},
  \bibinfo{person}{Brandon Houghton}, \bibinfo{person}{Kenny Hsu}, \bibinfo{person}{Shengli Hu}, \bibinfo{person}{Xin Hu}, \bibinfo{person}{Joost Huizinga}, \bibinfo{person}{Shantanu Jain}, \bibinfo{person}{Shawn Jain}, \bibinfo{person}{Joanne Jang}, \bibinfo{person}{Angela Jiang}, \bibinfo{person}{Roger Jiang}, \bibinfo{person}{Haozhun Jin}, \bibinfo{person}{Denny Jin}, \bibinfo{person}{Shino Jomoto}, \bibinfo{person}{Billie Jonn}, \bibinfo{person}{Heewoo Jun}, \bibinfo{person}{Tomer Kaftan}, \bibinfo{person}{Łukasz Kaiser}, \bibinfo{person}{Ali Kamali}, \bibinfo{person}{Ingmar Kanitscheider}, \bibinfo{person}{Nitish~Shirish Keskar}, \bibinfo{person}{Tabarak Khan}, \bibinfo{person}{Logan Kilpatrick}, \bibinfo{person}{Jong~Wook Kim}, \bibinfo{person}{Christina Kim}, \bibinfo{person}{Yongjik Kim}, \bibinfo{person}{Jan~Hendrik Kirchner}, \bibinfo{person}{Jamie Kiros}, \bibinfo{person}{Matt Knight}, \bibinfo{person}{Daniel Kokotajlo}, \bibinfo{person}{Łukasz Kondraciuk}, \bibinfo{person}{Andrew Kondrich},
  \bibinfo{person}{Aris Konstantinidis}, \bibinfo{person}{Kyle Kosic}, \bibinfo{person}{Gretchen Krueger}, \bibinfo{person}{Vishal Kuo}, \bibinfo{person}{Michael Lampe}, \bibinfo{person}{Ikai Lan}, \bibinfo{person}{Teddy Lee}, \bibinfo{person}{Jan Leike}, \bibinfo{person}{Jade Leung}, \bibinfo{person}{Daniel Levy}, \bibinfo{person}{Chak~Ming Li}, \bibinfo{person}{Rachel Lim}, \bibinfo{person}{Molly Lin}, \bibinfo{person}{Stephanie Lin}, \bibinfo{person}{Mateusz Litwin}, \bibinfo{person}{Theresa Lopez}, \bibinfo{person}{Ryan Lowe}, \bibinfo{person}{Patricia Lue}, \bibinfo{person}{Anna Makanju}, \bibinfo{person}{Kim Malfacini}, \bibinfo{person}{Sam Manning}, \bibinfo{person}{Todor Markov}, \bibinfo{person}{Yaniv Markovski}, \bibinfo{person}{Bianca Martin}, \bibinfo{person}{Katie Mayer}, \bibinfo{person}{Andrew Mayne}, \bibinfo{person}{Bob McGrew}, \bibinfo{person}{Scott~Mayer McKinney}, \bibinfo{person}{Christine McLeavey}, \bibinfo{person}{Paul McMillan}, \bibinfo{person}{Jake McNeil}, \bibinfo{person}{David
  Medina}, \bibinfo{person}{Aalok Mehta}, \bibinfo{person}{Jacob Menick}, \bibinfo{person}{Luke Metz}, \bibinfo{person}{Andrey Mishchenko}, \bibinfo{person}{Pamela Mishkin}, \bibinfo{person}{Vinnie Monaco}, \bibinfo{person}{Evan Morikawa}, \bibinfo{person}{Daniel Mossing}, \bibinfo{person}{Tong Mu}, \bibinfo{person}{Mira Murati}, \bibinfo{person}{Oleg Murk}, \bibinfo{person}{David Mély}, \bibinfo{person}{Ashvin Nair}, \bibinfo{person}{Reiichiro Nakano}, \bibinfo{person}{Rajeev Nayak}, \bibinfo{person}{Arvind Neelakantan}, \bibinfo{person}{Richard Ngo}, \bibinfo{person}{Hyeonwoo Noh}, \bibinfo{person}{Long Ouyang}, \bibinfo{person}{Cullen O'Keefe}, \bibinfo{person}{Jakub Pachocki}, \bibinfo{person}{Alex Paino}, \bibinfo{person}{Joe Palermo}, \bibinfo{person}{Ashley Pantuliano}, \bibinfo{person}{Giambattista Parascandolo}, \bibinfo{person}{Joel Parish}, \bibinfo{person}{Emy Parparita}, \bibinfo{person}{Alex Passos}, \bibinfo{person}{Mikhail Pavlov}, \bibinfo{person}{Andrew Peng}, \bibinfo{person}{Adam
  Perelman}, \bibinfo{person}{Filipe de Avila Belbute~Peres}, \bibinfo{person}{Michael Petrov}, \bibinfo{person}{Henrique~Ponde de Oliveira~Pinto}, \bibinfo{person}{Michael}, \bibinfo{person}{Pokorny}, \bibinfo{person}{Michelle Pokrass}, \bibinfo{person}{Vitchyr~H. Pong}, \bibinfo{person}{Tolly Powell}, \bibinfo{person}{Alethea Power}, \bibinfo{person}{Boris Power}, \bibinfo{person}{Elizabeth Proehl}, \bibinfo{person}{Raul Puri}, \bibinfo{person}{Alec Radford}, \bibinfo{person}{Jack Rae}, \bibinfo{person}{Aditya Ramesh}, \bibinfo{person}{Cameron Raymond}, \bibinfo{person}{Francis Real}, \bibinfo{person}{Kendra Rimbach}, \bibinfo{person}{Carl Ross}, \bibinfo{person}{Bob Rotsted}, \bibinfo{person}{Henri Roussez}, \bibinfo{person}{Nick Ryder}, \bibinfo{person}{Mario Saltarelli}, \bibinfo{person}{Ted Sanders}, \bibinfo{person}{Shibani Santurkar}, \bibinfo{person}{Girish Sastry}, \bibinfo{person}{Heather Schmidt}, \bibinfo{person}{David Schnurr}, \bibinfo{person}{John Schulman}, \bibinfo{person}{Daniel Selsam},
  \bibinfo{person}{Kyla Sheppard}, \bibinfo{person}{Toki Sherbakov}, \bibinfo{person}{Jessica Shieh}, \bibinfo{person}{Sarah Shoker}, \bibinfo{person}{Pranav Shyam}, \bibinfo{person}{Szymon Sidor}, \bibinfo{person}{Eric Sigler}, \bibinfo{person}{Maddie Simens}, \bibinfo{person}{Jordan Sitkin}, \bibinfo{person}{Katarina Slama}, \bibinfo{person}{Ian Sohl}, \bibinfo{person}{Benjamin Sokolowsky}, \bibinfo{person}{Yang Song}, \bibinfo{person}{Natalie Staudacher}, \bibinfo{person}{Felipe~Petroski Such}, \bibinfo{person}{Natalie Summers}, \bibinfo{person}{Ilya Sutskever}, \bibinfo{person}{Jie Tang}, \bibinfo{person}{Nikolas Tezak}, \bibinfo{person}{Madeleine~B. Thompson}, \bibinfo{person}{Phil Tillet}, \bibinfo{person}{Amin Tootoonchian}, \bibinfo{person}{Elizabeth Tseng}, \bibinfo{person}{Preston Tuggle}, \bibinfo{person}{Nick Turley}, \bibinfo{person}{Jerry Tworek}, \bibinfo{person}{Juan Felipe~Cerón Uribe}, \bibinfo{person}{Andrea Vallone}, \bibinfo{person}{Arun Vijayvergiya}, \bibinfo{person}{Chelsea Voss},
  \bibinfo{person}{Carroll Wainwright}, \bibinfo{person}{Justin~Jay Wang}, \bibinfo{person}{Alvin Wang}, \bibinfo{person}{Ben Wang}, \bibinfo{person}{Jonathan Ward}, \bibinfo{person}{Jason Wei}, \bibinfo{person}{CJ Weinmann}, \bibinfo{person}{Akila Welihinda}, \bibinfo{person}{Peter Welinder}, \bibinfo{person}{Jiayi Weng}, \bibinfo{person}{Lilian Weng}, \bibinfo{person}{Matt Wiethoff}, \bibinfo{person}{Dave Willner}, \bibinfo{person}{Clemens Winter}, \bibinfo{person}{Samuel Wolrich}, \bibinfo{person}{Hannah Wong}, \bibinfo{person}{Lauren Workman}, \bibinfo{person}{Sherwin Wu}, \bibinfo{person}{Jeff Wu}, \bibinfo{person}{Michael Wu}, \bibinfo{person}{Kai Xiao}, \bibinfo{person}{Tao Xu}, \bibinfo{person}{Sarah Yoo}, \bibinfo{person}{Kevin Yu}, \bibinfo{person}{Qiming Yuan}, \bibinfo{person}{Wojciech Zaremba}, \bibinfo{person}{Rowan Zellers}, \bibinfo{person}{Chong Zhang}, \bibinfo{person}{Marvin Zhang}, \bibinfo{person}{Shengjia Zhao}, \bibinfo{person}{Tianhao Zheng}, \bibinfo{person}{Juntang Zhuang},
  \bibinfo{person}{William Zhuk}, {and} \bibinfo{person}{Barret Zoph}.} \bibinfo{year}{2024}\natexlab{}.
\newblock \bibinfo{title}{GPT-4 Technical Report}.
\newblock
\newblock
\showeprint[arxiv]{2303.08774}~[cs.CL]
\urldef\tempurl%
\url{https://arxiv.org/abs/2303.08774}
\showURL{%
\tempurl}


\bibitem[Pham et~al\mbox{.}(2021)]%
        {pham2020problems}
\bibfield{author}{\bibinfo{person}{Hung~Viet Pham}, \bibinfo{person}{Shangshu Qian}, \bibinfo{person}{Jiannan Wang}, \bibinfo{person}{Thibaud Lutellier}, \bibinfo{person}{Jonathan Rosenthal}, \bibinfo{person}{Lin Tan}, \bibinfo{person}{Yaoliang Yu}, {and} \bibinfo{person}{Nachiappan Nagappan}.} \bibinfo{year}{2021}\natexlab{}.
\newblock \showarticletitle{Problems and opportunities in training deep learning software systems: an analysis of variance}. In \bibinfo{booktitle}{\emph{Proceedings of the 35th IEEE/ACM International Conference on Automated Software Engineering}} (Virtual Event, Australia) \emph{(\bibinfo{series}{ASE '20})}. \bibinfo{publisher}{Association for Computing Machinery}, \bibinfo{address}{New York, NY, USA}, \bibinfo{pages}{771–783}.
\newblock
\showISBNx{9781450367684}
\urldef\tempurl%
\url{https://doi.org/10.1145/3324884.3416545}
\showDOI{\tempurl}


\bibitem[Poesia et~al\mbox{.}(2022)]%
        {poesia2022synchromesh}
\bibfield{author}{\bibinfo{person}{Gabriel Poesia}, \bibinfo{person}{Oleksandr Polozov}, \bibinfo{person}{Vu Le}, \bibinfo{person}{Ashish Tiwari}, \bibinfo{person}{Gustavo Soares}, \bibinfo{person}{Christopher Meek}, {and} \bibinfo{person}{Sumit Gulwani}.} \bibinfo{year}{2022}\natexlab{}.
\newblock \bibinfo{title}{Synchromesh: Reliable code generation from pre-trained language models}.
\newblock
\newblock
\showeprint[arxiv]{2201.11227}~[cs.LG]
\urldef\tempurl%
\url{https://arxiv.org/abs/2201.11227}
\showURL{%
\tempurl}


\bibitem[Puigcerver et~al\mbox{.}(2024)]%
        {puigcerver2023sparse}
\bibfield{author}{\bibinfo{person}{Joan Puigcerver}, \bibinfo{person}{Carlos Riquelme}, \bibinfo{person}{Basil Mustafa}, {and} \bibinfo{person}{Neil Houlsby}.} \bibinfo{year}{2024}\natexlab{}.
\newblock \bibinfo{title}{From Sparse to Soft Mixtures of Experts}.
\newblock
\newblock
\showeprint[arxiv]{2308.00951}~[cs.LG]
\urldef\tempurl%
\url{https://arxiv.org/abs/2308.00951}
\showURL{%
\tempurl}


\bibitem[Radford and Narasimhan(2018)]%
        {radford2018improving}
\bibfield{author}{\bibinfo{person}{Alec Radford} {and} \bibinfo{person}{Karthik Narasimhan}.} \bibinfo{year}{2018}\natexlab{}.
\newblock \bibinfo{title}{Improving Language Understanding by Generative Pre-Training}.
\newblock
\newblock
\urldef\tempurl%
\url{https://api.semanticscholar.org/CorpusID:49313245}
\showURL{%
\tempurl}


\bibitem[Shaw et~al\mbox{.}(1975)]%
        {shaw1975inferring}
\bibfield{author}{\bibinfo{person}{David~E. Shaw}, \bibinfo{person}{William~R. Swartout}, {and} \bibinfo{person}{C.~Cordell Green}.} \bibinfo{year}{1975}\natexlab{}.
\newblock \showarticletitle{Inferring LISP programs from examples}. In \bibinfo{booktitle}{\emph{Proceedings of the 4th International Joint Conference on Artificial Intelligence - Volume 1}} (Tblisi, USSR) \emph{(\bibinfo{series}{IJCAI'75})}. \bibinfo{publisher}{Morgan Kaufmann Publishers Inc.}, \bibinfo{address}{San Francisco, CA, USA}, \bibinfo{pages}{260–267}.
\newblock


\bibitem[Smith(1975)]%
        {smith1975pygmalion}
\bibfield{author}{\bibinfo{person}{David~Canfield Smith}.} \bibinfo{year}{1975}\natexlab{}.
\newblock \bibinfo{title}{Pygmalion: a creative programming environment.}
\newblock
\newblock


\bibitem[Summers(1977)]%
        {summers1977methodology}
\bibfield{author}{\bibinfo{person}{Phillip~D Summers}.} \bibinfo{year}{1977}\natexlab{}.
\newblock \showarticletitle{A methodology for LISP program construction from examples}.
\newblock \bibinfo{journal}{\emph{Journal of the ACM (JACM)}} \bibinfo{volume}{24}, \bibinfo{number}{1} (\bibinfo{year}{1977}), \bibinfo{pages}{161--175}.
\newblock


\bibitem[Sun et~al\mbox{.}(2018)]%
        {sun2019grammar}
\bibfield{author}{\bibinfo{person}{Zeyu Sun}, \bibinfo{person}{Qihao Zhu}, \bibinfo{person}{Lili Mou}, \bibinfo{person}{Yingfei Xiong}, \bibinfo{person}{Ge Li}, {and} \bibinfo{person}{Lu Zhang}.} \bibinfo{year}{2018}\natexlab{}.
\newblock \bibinfo{title}{A Grammar-Based Structural CNN Decoder for Code Generation}.
\newblock
\newblock
\showeprint[arxiv]{1811.06837}~[cs.LG]
\urldef\tempurl%
\url{https://arxiv.org/abs/1811.06837}
\showURL{%
\tempurl}


\bibitem[Surameery and Shakor(2023)]%
        {surameery2023use}
\bibfield{author}{\bibinfo{person}{Nigar M~Shafiq Surameery} {and} \bibinfo{person}{Mohammed~Y Shakor}.} \bibinfo{year}{2023}\natexlab{}.
\newblock \showarticletitle{Use chat gpt to solve programming bugs}.
\newblock \bibinfo{journal}{\emph{International Journal of Information Technology \& Computer Engineering (IJITC) ISSN: 2455-5290}} \bibinfo{volume}{3}, \bibinfo{number}{01} (\bibinfo{year}{2023}), \bibinfo{pages}{17--22}.
\newblock


\bibitem[Svyatkovskiy et~al\mbox{.}(2020)]%
        {svyatkovskiy2020intellicode}
\bibfield{author}{\bibinfo{person}{Alexey Svyatkovskiy}, \bibinfo{person}{Shao~Kun Deng}, \bibinfo{person}{Shengyu Fu}, {and} \bibinfo{person}{Neel Sundaresan}.} \bibinfo{year}{2020}\natexlab{}.
\newblock \showarticletitle{IntelliCode compose: code generation using transformer}. In \bibinfo{booktitle}{\emph{Proceedings of the 28th ACM Joint Meeting on European Software Engineering Conference and Symposium on the Foundations of Software Engineering}} (Virtual Event, USA) \emph{(\bibinfo{series}{ESEC/FSE 2020})}. \bibinfo{publisher}{Association for Computing Machinery}, \bibinfo{address}{New York, NY, USA}, \bibinfo{pages}{1433–1443}.
\newblock
\showISBNx{9781450370431}
\urldef\tempurl%
\url{https://doi.org/10.1145/3368089.3417058}
\showDOI{\tempurl}


\bibitem[Vaithilingam et~al\mbox{.}(2022)]%
        {vaithilingam2022expectation}
\bibfield{author}{\bibinfo{person}{Priyan Vaithilingam}, \bibinfo{person}{Tianyi Zhang}, {and} \bibinfo{person}{Elena~L. Glassman}.} \bibinfo{year}{2022}\natexlab{}.
\newblock \showarticletitle{Expectation vs. Experience: Evaluating the Usability of Code Generation Tools Powered by Large Language Models}. In \bibinfo{booktitle}{\emph{Extended Abstracts of the 2022 CHI Conference on Human Factors in Computing Systems}} (New Orleans, LA, USA) \emph{(\bibinfo{series}{CHI EA '22})}. \bibinfo{publisher}{Association for Computing Machinery}, \bibinfo{address}{New York, NY, USA}, Article \bibinfo{articleno}{332}, \bibinfo{numpages}{7}~pages.
\newblock
\showISBNx{9781450391566}
\urldef\tempurl%
\url{https://doi.org/10.1145/3491101.3519665}
\showDOI{\tempurl}


\bibitem[Wang et~al\mbox{.}(2024)]%
        {wang2024software}
\bibfield{author}{\bibinfo{person}{Junjie Wang}, \bibinfo{person}{Yuchao Huang}, \bibinfo{person}{Chunyang Chen}, \bibinfo{person}{Zhe Liu}, \bibinfo{person}{Song Wang}, {and} \bibinfo{person}{Qing Wang}.} \bibinfo{year}{2024}\natexlab{}.
\newblock \showarticletitle{Software Testing With Large Language Models: Survey, Landscape, and Vision}.
\newblock \bibinfo{journal}{\emph{IEEE Trans. Softw. Eng.}} \bibinfo{volume}{50}, \bibinfo{number}{4} (\bibinfo{date}{Feb.} \bibinfo{year}{2024}), \bibinfo{pages}{911–936}.
\newblock
\showISSN{0098-5589}
\urldef\tempurl%
\url{https://doi.org/10.1109/TSE.2024.3368208}
\showDOI{\tempurl}


\bibitem[Wei et~al\mbox{.}(2019)]%
        {wei2019code}
\bibfield{author}{\bibinfo{person}{Bolin Wei}, \bibinfo{person}{Ge Li}, \bibinfo{person}{Xin Xia}, \bibinfo{person}{Zhiyi Fu}, {and} \bibinfo{person}{Zhi Jin}.} \bibinfo{year}{2019}\natexlab{}.
\newblock \bibinfo{title}{Code Generation as a Dual Task of Code Summarization}.
\newblock
\newblock
\showeprint[arxiv]{1910.05923}~[cs.LG]
\urldef\tempurl%
\url{https://arxiv.org/abs/1910.05923}
\showURL{%
\tempurl}


\bibitem[Wu et~al\mbox{.}(2020)]%
        {wu2020deep}
\bibfield{author}{\bibinfo{person}{Xiongfei Wu}, \bibinfo{person}{Liangyu Qin}, \bibinfo{person}{Bing Yu}, \bibinfo{person}{Xiaofei Xie}, \bibinfo{person}{Lei Ma}, \bibinfo{person}{Yinxing Xue}, \bibinfo{person}{Yang Liu}, {and} \bibinfo{person}{Jianjun Zhao}.} \bibinfo{year}{2020}\natexlab{}.
\newblock \showarticletitle{How are Deep Learning Models Similar? An Empirical Study on Clone Analysis of Deep Learning Software}. In \bibinfo{booktitle}{\emph{Proceedings of the 28th International Conference on Program Comprehension}} (Seoul, Republic of Korea) \emph{(\bibinfo{series}{ICPC '20})}. \bibinfo{publisher}{Association for Computing Machinery}, \bibinfo{address}{New York, NY, USA}, \bibinfo{pages}{172–183}.
\newblock
\showISBNx{9781450379588}
\urldef\tempurl%
\url{https://doi.org/10.1145/3387904.3389254}
\showDOI{\tempurl}


\bibitem[Xu et~al\mbox{.}(2022)]%
        {xu2022ide}
\bibfield{author}{\bibinfo{person}{Frank~F Xu}, \bibinfo{person}{Bogdan Vasilescu}, {and} \bibinfo{person}{Graham Neubig}.} \bibinfo{year}{2022}\natexlab{}.
\newblock \showarticletitle{In-ide code generation from natural language: Promise and challenges}.
\newblock \bibinfo{journal}{\emph{ACM Transactions on Software Engineering and Methodology (TOSEM)}} \bibinfo{volume}{31}, \bibinfo{number}{2} (\bibinfo{year}{2022}), \bibinfo{pages}{1--47}.
\newblock


\bibitem[Yetiştiren et~al\mbox{.}(2023)]%
        {yeticstiren2023evaluating}
\bibfield{author}{\bibinfo{person}{Burak Yetiştiren}, \bibinfo{person}{Işık Özsoy}, \bibinfo{person}{Miray Ayerdem}, {and} \bibinfo{person}{Eray Tüzün}.} \bibinfo{year}{2023}\natexlab{}.
\newblock \bibinfo{title}{Evaluating the Code Quality of AI-Assisted Code Generation Tools: An Empirical Study on GitHub Copilot, Amazon CodeWhisperer, and ChatGPT}.
\newblock
\newblock
\showeprint[arxiv]{2304.10778}~[cs.SE]
\urldef\tempurl%
\url{https://arxiv.org/abs/2304.10778}
\showURL{%
\tempurl}


\bibitem[Yin and Neubig(2017)]%
        {yin2017syntactic}
\bibfield{author}{\bibinfo{person}{Pengcheng Yin} {and} \bibinfo{person}{Graham Neubig}.} \bibinfo{year}{2017}\natexlab{}.
\newblock \bibinfo{title}{A Syntactic Neural Model for General-Purpose Code Generation}.
\newblock
\newblock
\showeprint[arxiv]{1704.01696}~[cs.CL]
\urldef\tempurl%
\url{https://arxiv.org/abs/1704.01696}
\showURL{%
\tempurl}


\bibitem[Yin and Neubig(2018)]%
        {yin2018tranx}
\bibfield{author}{\bibinfo{person}{Pengcheng Yin} {and} \bibinfo{person}{Graham Neubig}.} \bibinfo{year}{2018}\natexlab{}.
\newblock \showarticletitle{TRANX: A Transition-based Neural Abstract Syntax Parser for Semantic Parsing and Code Generation}. In \bibinfo{booktitle}{\emph{Proceedings of the 2018 Conference on Empirical Methods in Natural Language Processing: System Demonstrations}}, \bibfield{editor}{\bibinfo{person}{Eduardo Blanco} {and} \bibinfo{person}{Wei Lu}} (Eds.). \bibinfo{publisher}{Association for Computational Linguistics}, \bibinfo{address}{Brussels, Belgium}, \bibinfo{pages}{7--12}.
\newblock
\urldef\tempurl%
\url{https://doi.org/10.18653/v1/D18-2002}
\showDOI{\tempurl}


\bibitem[Yu et~al\mbox{.}(2024)]%
        {yu2024codereval}
\bibfield{author}{\bibinfo{person}{Hao Yu}, \bibinfo{person}{Bo Shen}, \bibinfo{person}{Dezhi Ran}, \bibinfo{person}{Jiaxin Zhang}, \bibinfo{person}{Qi Zhang}, \bibinfo{person}{Yuchi Ma}, \bibinfo{person}{Guangtai Liang}, \bibinfo{person}{Ying Li}, \bibinfo{person}{Qianxiang Wang}, {and} \bibinfo{person}{Tao Xie}.} \bibinfo{year}{2024}\natexlab{}.
\newblock \showarticletitle{CoderEval: A Benchmark of Pragmatic Code Generation with Generative Pre-trained Models}. In \bibinfo{booktitle}{\emph{Proceedings of the IEEE/ACM 46th International Conference on Software Engineering}} (Lisbon, Portugal) \emph{(\bibinfo{series}{ICSE '24})}. \bibinfo{publisher}{Association for Computing Machinery}, \bibinfo{address}{New York, NY, USA}, Article \bibinfo{articleno}{37}, \bibinfo{numpages}{12}~pages.
\newblock
\showISBNx{9798400702174}
\urldef\tempurl%
\url{https://doi.org/10.1145/3597503.3623316}
\showDOI{\tempurl}


\bibitem[Zan et~al\mbox{.}(2022)]%
        {zan2022cert}
\bibfield{author}{\bibinfo{person}{Daoguang Zan}, \bibinfo{person}{Bei Chen}, \bibinfo{person}{Dejian Yang}, \bibinfo{person}{Zeqi Lin}, \bibinfo{person}{Minsu Kim}, \bibinfo{person}{Bei Guan}, \bibinfo{person}{Yongji Wang}, \bibinfo{person}{Weizhu Chen}, {and} \bibinfo{person}{Jian-Guang Lou}.} \bibinfo{year}{2022}\natexlab{}.
\newblock \bibinfo{title}{CERT: Continual Pre-Training on Sketches for Library-Oriented Code Generation}.
\newblock
\newblock
\showeprint[arxiv]{2206.06888}~[cs.SE]
\urldef\tempurl%
\url{https://arxiv.org/abs/2206.06888}
\showURL{%
\tempurl}


\end{thebibliography}

\appendix









\end{document}